\documentclass[11pt,hyper,a4paper]{article}
\usepackage{jheppub}
\usepackage{amsmath,amssymb,amsfonts,amscd,bm, amsthm, slashed, mathrsfs, bm}
\usepackage{graphicx, wrapfig}
\usepackage{multirow}
\usepackage{verbatim}
\usepackage[title]{appendix}
\usepackage{fancybox}
\usepackage[utf8]{inputenc}
\usepackage{arydshln}
\usepackage{dsfont}

\usepackage{url}
\usepackage{float}

\usepackage[normalem]{ulem}
\usepackage{graphicx}

\usepackage{color}
\usepackage[usenames,dvipsnames,svgnames,table]{xcolor}
\definecolor{darkblue}{cmyk}{0.9,0.9,0,0}
\hypersetup{colorlinks=false, linkcolor=darkblue, linkcolor=darkblue, citecolor=darkblue}

\usepackage{bbm}
\usepackage{comment}
\usepackage{enumitem}
\usepackage{float}
\usepackage{empheq}
\usepackage{enumerate}

\title{3d-3d correspondence for knot complements with finite and large $N$}

\author{Hee-Joong Chung}

\affiliation{Department of Science Education, Jeju National University, Jeju, 63243, Republic of Korea}

\abstract{For $G=SU(N)$ at finite and large $N$, with a totally symmetric representation, we realize the homological block $F_K$ for a knot complement $S^3 \backslash K$, given in the form of the inverted Habiro series, as a half-index of a 3d $\mathcal{N}=2$ theory $T[M_3]$ by studying some examples, which we expect to extend to general knots.
From the half-index expression, it is also possible to realize the colored HOMFLY-PT polynomial by taking a certain set of poles.
Through the half-index realization, we describe a method for obtaining the $G=SU(N)$ homological block and its $a$-deformed version for $S^3 \backslash K$ from a Habiro series expression for the colored HOMFLY-PT polynomial.
We also discuss some properties of partition functions for arbitrary $N$.
}

\begin{document}

\maketitle


\section{Introduction}

The 3d-3d correspondence states that $G_\mathbb{C}$ Chern-Simons theory on a 3-manifold $M_3$ is related to a 3d $\mathcal{N}=2$ Chern-Simons matter theory $T[M_3; G]$ \cite{Dimofte-Gaiotto-Gukov, Dimofte-Gukov-Hollands, Terashima-Yamazaki}.
In the early development of the correspondence, the 3d $\mathcal{N}=2$ theory $T[M_3; G]$ was constructed by using the ideal tetrahedra as building blocks, but it didn't capture the abelian branch that is present in Chern-Simons theory \cite{Dimofte-Gaiotto-Gukov, Chung-Dimofte-Gukov-Sulkowski}, in particular for a knot complement $S^3 \backslash K$.		\\

Meanwhile, homological blocks were introduced in \cite{Gukov-Putrov-Vafa, Gukov-Pei-Putrov-Vafa, Gukov-Manolescu}.
They can be regarded as the contributions from abelian flat connections to the partition function of the analytically continued Chern-Simons theory, and they also encode the contributions from non-abelian flat connections \cite{Gukov-Marino-Putrov, Chung-resurg}.
Therefore, the homological block would be a natural quantity that arises from the 3d $\mathcal{N}=2$ theory $T[M_3]$ that captures all flat connections, and it is realized as a half-index \cite{GGP-walls, Sugiyama-Yoshida, Dimofte-Gaiotto-Paquette} of $T[M_3]$ with appropriate boundary conditions preserving 2d $\mathcal{N}=(0,2)$ supersymmetries.	\\

For a knot complement $S^3 \backslash K$, there are two types of expansions for homological blocks, which are the balanced expansion and the positive (or negative) expansion.
The balanced expansion of the homological block $F_K(x,q)$ for $G=SU(2)$ is a Weyl-invariant expression under the Weyl action $x \leftrightarrow x^{-1}$, and the positive (resp. negative) expansion is obtained from the series expansion of $F_K(x,q)$ in $x$ (resp. in $x^{-1}$).
The positive expansion admits a quiver form, often called a fermionic sum, and it can be realized as a half-index of a 3d $\mathcal{N}=2$ theory $T[M_3]$ with appropriate boundary conditions \cite{Chung-3d3dpe}.
The inverted Habiro series \cite{Park-inverted} for a homological block is manifestly Weyl invariant and serves as a closed form expression for the balanced expansion.
The half-index realization of the homological block in the form of the inverted Habiro series was obtained in \cite{Chung-ab} for $G=SU(2)$, which leads to $T[S^3 \backslash K, SU(2)]$ that captures all flat connections.	\\

In this work, we extend \cite{Chung-ab} to the case of the gauge group $G=SU(N)$ with general $N$ and a totally symmetric representation of $G$.
That is, we realize the $G=SU(N)$ homological block as a half-index of a 3d $\mathcal{N}=2$ theory with suitable boundary conditions and an appropriate choice of integration contour by working out several examples, which we expect to extend to general knots.
The $N$-dependence of the homological block from the inverted Habiro series for $G=SU(N)$ can be packaged as $a=q^N$.
Therefore, we also obtain the $a$-deformed version of the inverted Habiro series and a half-index realization of it. 
A half-index realization of the homological block leads to a 3d $\mathcal{N}=2$ theory that captures all branches of flat connections.
We also see that, for $G=SU(N)$, the contour chosen in the integral expression for the half-index to obtain the homological block is naturally associated with the critical point for the abelian branch, whereas in the $a$-deformed case the branches are not distinguished as abelian or non-abelian but are instead on equal footing.

The homological block and the colored HOMFLY-PT polynomial are related by choosing different sets of poles in the half-index realization, and we can obtain the inverted Habiro series expression of the homological block for a given colored HOMFLY-PT polynomial in the form of the Habiro series.
By using this method, we also obtain the homological block for the mirror $S^3 \backslash m(K)$, which provides the extension of the homological block from $|q|<1$ to $|q|>1$.
In addition, we propose a way to obtain the homological block for the supergroup $GL(L|M)$ by using the relation between the reduced colored $U_q(\mathfrak{gl}_{L|M})$ Reshetikhin-Turaev invariants of knots and the reduced colored HOMFLY-PT polynomials at $a=q^{L-M}$.

Also, we compute the partition functions of $T[M_3]$ for finite and large $N$.
In particular, for the $S^3_b$ partition function, it provides an $a$-deformed state integral model for a knot complement $S^3 \backslash K$.
From the factorization of the partition functions, we relate the anti-half-index for a knot to the half-index for its mirror knot.

We also consider the partition functions at and near $a=q^N$.
Upon taking a residue at $(a,\tilde{a})=(q^N, \tilde{q}^{\tilde{N}})$, which is interpreted as Higgsing $U(1)_a$ symmetry, the resulting partition functions contain contributions only from the non-abelian branches.
Also, when $a=q^N$, we see that the ratio of the contribution from the abelian branch to that from the non-abelian branch becomes zero.
In addition, if taking the $\epsilon$-expansion of the partition functions with $a=q^{N+\epsilon}$, the contribution from the non-abelian branch appears at leading order, while that from the abelian branch appears at next-to-leading order.	
These are consistent with the general form of the partition function of complex Chern-Simons theory.	\\

In section 2, we discuss the half-index realization of the homological block for $G=SU(N)$ and a totally symmetric representation, and its $a$-deformed version.
While doing so, we propose a way to obtain the $G=SU(N)$ homological block in the form of an inverted Habiro series, given a HOMFLY-PT polynomial with a totally symmetric representation in the form of a Habiro series.
We study the critical points and the associated contours for $a=q^N$ and generic $a$.
We also propose a way to obtain the homological block for $GL(L|M)$ via the HOMFLY-PT polynomial with a totally symmetric representation.
In section 3, we compute the partition functions and the twisted indices, and discuss their properties.
The handle-gluing and fibering operators of the twisted indices on a degree-$p$ $S^1$-bundle over a genus-$g$ Riemann surface, $\mathcal{M}_{g,p}$, for several knot complements are given in Appendix A.


\section{Homological blocks for knot complements and $T[M_3]$ with arbitrary $N$}
\label{arbn}

We consider the homological block for a knot complement in the form of the inverted Habiro series for $G=SU(N)$ and a totally symmetric representation.
The Habiro series expression of the colored HOMFLY-PT polynomial for the totally symmetric representation of $G=SU(N)$ with $n-1$ boxes is given by
\begin{align}
P_K(n, a, q) = \sum_{k =0}^{\infty} \mathfrak{a}_k(K, a, q) \begin{bmatrix} n-1 \\ k \end{bmatrix} \prod_{j=1}^{k} (a^{\frac{1}{2}} q^{\frac{n+j-2}{2}} - a^{-\frac{1}{2}} q^{-\frac{n+j-2}{2}}) 
(a^{\frac{1}{2}} q^{\frac{j-2}{2}} - a^{-\frac{1}{2}} q^{-\frac{j-2}{2}}) 
\label{homfly}
\end{align}
where the Habiro coefficients $\mathfrak{a}_k(K;a,q)$, $k=0,1, \ldots$, are Laurent polynomials in $a=q^N$ and $q$, and they satisfy the $q$-difference equation given by the quantum $C$-polynomial \cite{Garoufalidis-Sun, Mironov-Morozov-C}.
Here, 
\begin{align}
[l] := \frac{q^{\frac{l}{2}} - q^{-\frac{l}{2}}}{q^{\frac{1}{2}} - q^{-\frac{1}{2}}}	\,	,	\quad
[l]! := \prod_{j=1}^l [j]	\,	,	\quad
\begin{bmatrix} l \\ m \end{bmatrix} := \frac{[l]!}{[m]! [l-m]!}	\,	.
\end{align}
For $G=SU(N)$ and the the totally symmetric representation, it was conjectured \cite{Park-inverted} that the homological block for a knot complement is given by
\begin{align}
F_K^{SU(N)}(x,q) = - \sum_{k=0}^{\infty} \alpha_{-k-1}(K; N, q) \frac{\frac{[-n] [-n+1] \cdots [-n+N-3]}{[N-2]!}}{\prod_{j=0}^{k}  (x^{\frac{1}{2}} q^{\frac{j}{2}} - x^{-\frac{1}{2}} q^{-\frac{j}{2}}) (x^{\frac{1}{2}} q^{\frac{(N-2)-j}{2}} - x^{-\frac{1}{2}} q^{-\frac{(N-2)-j}{2}})  }
\label{hbn}
\end{align}
where $n = \frac{\log x}{\log q}$.
The inverted Habiro coefficients $\alpha_{-k-1}(K; N, q)$, $k=0,1,\ldots$, are Laurent series in $q$, and are given by extending $\mathfrak{a}_k(K;a,q)$ to negative values of $k$.
The homological block \eqref{hbn} is a normalized or reduced version, and the unnormalized or unreduced version is obtained by multiplying it by the homological block for the unknot.\footnote{For $G=SU(N)$ and the totally symmetric representation, the homological block for the unknot is given by $e^{-\frac{1}{2\hbar} \log(q^{-1}x) \log a} (q^{-1}x)^{\frac{1}{2}} \frac{(a;q)_\infty (x;q)_\infty}{(q^{-1}xa;q)_\infty (q;q)_\infty}$ \cite{EGGKPSS}.}
The homological block \eqref{hbn} can also be expressed as 
\begin{align}
F_K^{SU(N)}(x,q) = \sum_{k=0}^{\infty} \alpha_{-k-1}(K; N, q) \frac{(-1)^k q^{\frac{1}{2}k(k+1)}}{(x;q)_{k+N-1} (x^{-1} q^{2-N};q)_{k+N-1}} 
\frac{(q^{k+1};q)_\infty}{(q;q)_\infty}
\frac{(q^{N-1};q)_\infty}{(q^{k+N-1};q)_\infty}	\,	,
\label{hbn1}
\end{align}
where $(w;q)_n = \prod_{j=0}^{n-1} (1-wq^j)$ is the $q$-Pochhammer symbol.
This expression is manifestly Weyl symmetric under $x \leftrightarrow q^{2-N} x^{-1}$.
Both the colored HOMFLY-PT polynomial \eqref{homfly} and homological block \eqref{hbn} are expected to be annihilated by the same quantum $\hat{A}$ and $\hat{B}$-polynomials \cite{EGGKPS, EGGKPSS}.


\subsection{Homological blocks and half-indices of $T[M_3]$}
\label{ssec:hbhi}

For concreteness, we begin with the case of the figure-eight knot where the inverted Habiro coefficients are $\alpha_{-k-1}(4_1;N,q)=1$ \cite{Park-inverted}.
We expect that the method described in this work and in \cite{Chung-ab} for obtaining the half-index realization of the homological block in the form of an inverted Habiro series for totally symmetric representations of $G=SU(N)$ with arbitrary $N$ can be extended to general knots.	\\

From \eqref{hbn1}, the homological block for the figure-eight knot is given by
\begin{align}
F_{4_1}^{SU(N)}(x,q) = \sum_{k=0}^{\infty} (-1)^k q^{\frac{1}{2}k(k+1)} 
\frac{(q^{k+1};q)_\infty}{(q;q)_\infty}
\frac{(q^{N-1};q)_\infty}{(q^{k+N-1};q)_\infty}
\frac{(q^{k+N-1} x;q)_\infty}{(x;q)_\infty} \frac{(q^{k+1} x^{-1} ;q)_\infty}{(q^{2-N} x^{-1} ;q)_\infty}	\,	.
\label{hb41n}
\end{align}
Since the $N$-dependence appears as $q^N$ in \eqref{hb41n}, we obtain the $a$-deformed homological block in closed form
\begin{align}
\hspace{-3mm}F_{4_1}(x,a,q) = \sum_{k=0}^{\infty} (-1)^k q^{\frac{1}{2}k(k+1)} 
\frac{(q^{k+1};q)_\infty}{(q;q)_\infty}
\frac{(q^{-1}a;q)_\infty}{(q^{k-1}a;q)_\infty}
\frac{(xaq^{k-1};q)_\infty}{(x;q)_\infty} \frac{(x^{-1} q^{k+1};q)_\infty}{(x^{-1} a^{-1}q^{2};q)_\infty}		\,	.
\label{hb41a}
\end{align}
The power series expansion of \eqref{hb41a} in $x$ agrees with that obtained in \cite{EGGKPSS}.
When $N=1$ or $a=q$, the homological block is trivial, \textit{i.e.} equal to 1, as expected.		\\

From the asymptotic limit of \eqref{hb41a}, the effective twisted superpotential is obtained as
\begin{align}
\begin{split}
\widetilde{\mathcal{W}}_{4_1}(z,x,a) &= \frac{1}{2} (\log z)^2 + \log(-1) \log z+ \text{Li}_2(zxa) - \text{Li}_2(x) + \text{Li}_2(x^{-1}z) \\
&\hspace{40mm}- \text{Li}_2(x^{-1}a^{-1}) + \text{Li}_2(z) - \text{Li}_2(za) + \text{Li}_2(a)	\,	.
\end{split}
\label{twspot41}
\end{align}
From the supersymmetric vacuum condition,
\begin{align}
1 = \exp \frac{\partial \widetilde{\mathcal{W}}(z,x,a)}{\partial \log z}	\,	,
\label{svac}
\end{align}
and the equations for the supersymmetric parameter space
\begin{align}
y = \exp \frac{\partial \widetilde{\mathcal{W}}(z,x,a)}{\partial \log x}	\,	,	\quad	
b = \exp \frac{\partial \widetilde{\mathcal{W}}(z,x,a)}{\partial \log a}	\,	,
\label{spar}
\end{align}
we obtain
\begin{align}
\begin{split}
&\hspace{-2mm} 0 = \widetilde{A}_{4_1}(x,y,a) = (x-1) (a x-1) 	\\
&\hspace{25mm}\times \big(a x^2 (a x-1)^2 y^3 
-(a^4 x^6-3 a^3 x^5+2 a^2 x^4+2 a^2 x^2-3 a x+1) y^2 	\\
&\hspace{47mm}+ a (x-1)(a^2 x^5 - 2 a^2 x^4 + 2 x - 1) y 
-a^2 (x-1)^2 x^2 \big)	\,	,
\end{split}	\label{aapoly41tw}	\\
\begin{split}
&\hspace{-2mm} 0=\widetilde{B}_{4_1}(a,b,x) = a (a-1) (x-1) (a x-1) 	\\
&\hspace{17mm}\times \big(
-x (a-1)^2 (a x-1)^2 b^3 +(a-1) (a x-1) (a^2 x^3-2 a (x+1) x+1 ) b^2 	\\
&\hspace{63mm}+a x (2 a^2 x^2 - (x^2+3 x+1)a + 2 ) b 
+x^2 a^2 \big)	\,	,
\end{split}	\label{xbpoly41tw}
\end{align}
where the last factors on the right-hand sides of \eqref{aapoly41tw} and \eqref{xbpoly41tw} are the classical $a$-deformed $A$-polynomial and $x$-deformed $B$-polynomial, respectively.
They become
\begin{align}
&-(x-1)^2 (y-1) \big(x^2 y^2 - (x^4-x^3-2 x^2-x+1) y +x^2 \big)	\,	,	\label{capoly41}	\\
&- (b-1) ( (a-1)^2 b+a )^2	\,	,	\label{cbpoly41}
\end{align}
when $a=1$ and $x=1$, respectively.
In \eqref{capoly41} and \eqref{cbpoly41}, $y-1$ and $b-1$ correspond to the abelian branch, and $x^2 y^2 - (x^4-x^3-2 x^2-x+1) y +x^2$ and $( (a-1)^2 b+a )^2$ correspond to the non-abelian branches.	\\

The homological block \eqref{hb41n} and its $a$-deformed version \eqref{hb41a} are annihilated by the quantum $\hat{A}$-polynomial for the figure-eight knot,
\begin{align}
&\hat{A}_{4_1}(\hat{x},\hat{y},a,q) = a q^3 \hat{x}^2 (a \hat{x} - 1) (a \hat{x}^2 - 1) (a q \hat{x} - 1) (a \hat{x}^2 - q) \hat{y}^3	\nonumber	\\
&\hspace{20mm} +\big(a^3 q^3 \hat{x}^5 - 2 a^2 q^3 \hat{x}^4 - a^2 (q-1)^2 (q+1) \hat{x}^3 + a (q-1)^2 (q+1) \hat{x}^2 + 2 a \hat{x} - 1 \big) \hat{y}^2		\nonumber	\\
&\hspace{20mm} +\big(a^2 q^3 \hat{x}^5 - 2 a^2 q^2 \hat{x}^4 - a q (q-1)^2 (q+1) \hat{x}^3 + a (q-1)^2 (q+1) \hat{x}^2 + 2 q^2 \hat{x} - q \big) \hat{y}		\nonumber	\\
&\hspace{20mm} -a^2 q \hat{x}^2 (\hat{x} - 1) (q \hat{x} - 1) (a q^2 \hat{x}^2 - 1) (a q^3 \hat{x}^2 - 1)	\,	.	
\label{qapoly41}
\end{align}
Here, $\hat{y}f(x)=f(qx)$ and $\hat{x}f(x) =xf(x)$ on functions of $x$, and $\hat{y}f(n)=f(n+1)$ and $\hat{x}f(n) =q^nf(n)$ on functions of $n$, with $\hat{y} \hat{x} = q \hat{x} \hat{y}$ in both cases.
The quantum $\hat{A}$-polynomial \eqref{qapoly41} is obtained by using the \textit{Mathematica} package \texttt{HolonomicFunctions} \cite{Koutschan:holofunctions}, which is used in other examples in this work.
The quantum $\hat{A}$-polynomial \eqref{qapoly41} agrees with the one obtained in \cite{FGS-superA}, and it annihilates the colored HOMFLY-PT polynomial \eqref{homfly} with $\mathfrak{a}_k(4_1,a,q)=1$.

The homological block \eqref{hb41n}, its $a$-deformed version \eqref{hb41a}, and the colored HOMFLY-PT polynomial \eqref{homfly} for the figure-eight knot are also annihilated by the quantum $\hat{B}$-polynomial,
\begin{align}
\begin{split}
&\hat{B}_{4_1}(\hat{a},\hat{b},x,q) = - x (\hat{a} - 1) (q\hat{a} - 1) (x \hat{a} - 1) (q x \hat{a} - 1) \hat{b}^3 	\\
&\hspace{27mm} + q (\hat{a}-1) \big(x^4 \hat{a}^3 - x^2 (3 x+2) \hat{a}^2 + x (2 x+3) \hat{a} - 1 \big) \hat{b}^2 	\\
&\hspace{27mm} + \big( x^2 (q+1) \hat{a}^2 - q (x^2+3 x+1)\hat{a} + q^2 + q \big) \hat{b} 	 + q x^2 \hat{a}^2	\,	.
\end{split}
\label{qbpoly41}
\end{align}
This agrees with the quantum $\hat{B}$-polynomial for the figure-eight knot in \cite{EGGKPSS}.
Here, $\hat{b}f(a)=f(qa)$ and $\hat{a} f(a) = a f(a)$ on functions of $a$, and $\hat{b}f(N)=f(N+1)$ and $\hat{a} f(N) = q^N f(N)$ on functions of $N$, with $\hat{b}\hat{a}=q\hat{a}\hat{b}$ in both cases.

The classical limit $q \rightarrow 1$ of the quantum $\hat{A}$-polynomial \eqref{qapoly41} gives
\begin{align}
\begin{split}
&A_{4_1}(x,y,a) = (a x^2-1)^2 
\big(a x^2 (a x-1)^2 y^3 -(a x-1) (a^3 x^5-2 a^2 x^4+2 a x-1)y^2 \\
&\hspace{45mm}+ a (x-1) (a^2 x^5-2 a^2 x^4+2 x-1) y -a^2 (x-1)^2 x^2 \big)	\,	,
\end{split}
\label{aapoly41}
\end{align}
and the second factor agrees with the classical $a$-deformed $A$-polynomial in \eqref{aapoly41tw} obtained from the effective twisted superpotential.
Similarly, in the limit $q\rightarrow 1$, the quantum $\hat{B}$-polynomial \eqref{qbpoly41} becomes the classical $x$-deformed $B$-polynomial
\begin{align}
\begin{split}
&B_{4_1}(a,b,x) = -(a-1)^2 x (a x-1)^2 b^3  
+ (a-1) (a x-1) (a^2 x^3-2 a (x+1) x+1) b^2 \\
&\hspace{25mm}+ a x (2 a^2 x^2 -(x^2+3 x+1)a + 2) b 
+a^2 x^2	\,	,
\end{split}
\label{bapoly41}
\end{align}
which also agrees with the one in \eqref{xbpoly41tw} obtained from the effective twisted superpotential.	\\

When $N \in \mathbb{Z}_{\geq 2}$, applying the operator $\hat{A}_{4_1}^{\text{nab}}(N;\hat{x},\hat{y},q)$, which corresponds to the non-abelian branch, to \eqref{hb41n} leads to an inhomogeneous equation.
This inhomogeneous term is annihilated by another operator $\hat{A}_{4_1}^{\text{ab}}(N;\hat{x},\hat{y},q)$ corresponding to the abelian branch.
For example, when $N=3$, 
\begin{align}
\begin{split}
&\hspace{-2mm}\hat{A}_{4_1}^{\text{nab}}(3;\hat{x},\hat{y},q) = 
q^3 \hat{x}^2 (q \hat{x}+1) (q^3 \hat{x} - 1) (q^2 \hat{x}^2 - 3 q \hat{x} + 1) \hat{y}^2	\\
&\hspace{25mm} -\big( q^{12} x^8 - 2 (q+1) q^{10} \hat{x}^7 - (q^2 - 3q + 1) q^8 \hat{x}^6 + (q+1)^3 q^6 \hat{x}^5 \\
&\hspace{45mm} - (q+1)^3 q^3 \hat{x}^3 + (q^2 - 3 q + 1) q^2 \hat{x}^2 + 2 (q+1) q \hat{x} - 1\big) \hat{y}	\\
&\hspace{25mm} +q^3 \hat{x}^2 (\hat{x} - 1) (q^2 \hat{x} + 1) (q^4 \hat{x}^2 - 3 q^2 \hat{x} + 1)	\,	,
\end{split}	\label{qapolynab3}	\\
&\hat{A}_{4_1}^{\text{ab}}(3;\hat{x},\hat{y},q) = (q \hat{x} + 1) (q^3 \hat{x}^2-1) \hat{y} - q^2 (q^3 \hat{x} + 1) (q^5 \hat{x}^2 - 1)	\,	.
\label{qapolyab3}	
\end{align}
Their classical limits are given by
\begin{align}
A_{4_1}^{\text{nab}}(3;x,y) &= (x^4-3 x^3+3 x-1) \big(x^2 y^2 - (x^4-x^3-2 x^2-x+1) y +x^2 \big)	\,	,	\\
A_{4_1}^{\text{ab}}(3;x,y) &= (x-1) (x+1)^2 (y-1)	\,	,
\end{align}
and indeed they correspond to the non-abelian and abelian branches, respectively.
We have also checked that $\hat{A}_{4_1}^{\text{ab}}(N;\hat{x},\hat{y},q) \hat{A}_{4_1}^{\text{nab}}(N;\hat{x},\hat{y},q)$ agrees with $\hat{A}_{4_1}(\hat{x},\hat{y},q^N,q) $ in \eqref{qapoly41} up to an overall factor for several values of $N$, and this relation should hold for general $N$.	\\

We can also obtain the quantum $\widehat{AB}$-ideal annihilating the homological block $F_K(x,a,q) $\cite{EGGKPSS}, 
\begin{align}
\widehat{AB}_K = \big\{ \hat{T} \in \mathbb{C} [ \hat{x}^{\pm1}, \hat{a}^{\pm1}, \hat{y}^{\pm1}, \hat{b}^{\pm1}, q^{\pm1}] \} \big| \, \hat{T} F_{K}(x,a,q) =0 \big\}	\,	,
\label{abid}
\end{align}
for the figure-eight knot.
With respect to the lexicographic order $\hat{a} \succ \hat{x}$, we have a Gr\"obner basis of the annihilating ideal $\{ \hat{D}_{4_1}, \hat{A}_{4_1}\}$ of the homological block \eqref{hb41a}\footnote{Though it is suspected that this Gr\"obner basis generates the whole ideal, it may generate only a subideal of \eqref{abid}.} where $\hat{A}_{4_1}$ is given by \eqref{qapoly41} and 
\begin{align}
\hat{D}_{4_1}(\hat{x}, \hat{a}, \hat{y}, \hat{b},q) = - (\hat{a}-q) (\hat{a} \hat{x}^2 - q ) \hat{b} + q \hat{a} \hat{x} \hat{y} - q \hat{a} \hat{x}	\,	.	\label{opd41}
\end{align}
For the lexicographic order $\hat{x} \succ \hat{a}$, a Gr\"obner basis of the annihilating ideal is given by $\{ \hat{D}_{4_1}, \hat{B}_{4_1}\}$, where $\hat{B}_{4_1}$ is given by \eqref{qbpoly41}.


\subsubsection*{Half-index realizations of homological blocks}

The homological block \eqref{hb41n} and \eqref{hb41a} can be obtained from the integral
\begin{align}
(q;q)_\infty \int \frac{dz}{2 \pi i z} \Upsilon_{4_1}(z,x,a,q)
\label{int41a}
\end{align}
with the integrand
\begin{align}
\Upsilon_{4_1}(z,x,a,q) = \frac{1}{(z^{-1};q)_\infty} \frac{(q^{-1}a;q)_\infty}{(q^{-1}az;q)_\infty} \frac{(q^{-1}axz;q)_\infty}{(x;q)_\infty} \frac{(q x^{-1} z;q)_\infty}{(q^2 a^{-1} x^{-1} ;q)_\infty}	\,	,	\label{integ41a}
\end{align}
by taking the poles $z=q^k$, $k=0,1, \ldots$, from $(z^{-1};q)_\infty^{-1}$ in \eqref{integ41a}.
Here,
\begin{align}
\theta(u;q) := (-q^{\frac{1}{2}} u;q)_\infty (-q^{\frac{1}{2}} u^{-1};q)_\infty
\end{align}
is the Jacobi theta function divided by $(q;q)_\infty$.

The 3d $\mathcal{N}=2$ theory whose half-index is \eqref{int41a} has field contents under $U(1)_z \times U(1)_x \times U(1)_a \times U(1)_R$ symmetries with boundary conditions: 
a $U(1)_z$ vector multiplet with Neumann boundary condition;
seven 3d chiral multiplets, $\Phi_{l=1, \ldots, 7}$, where
$\Phi_{l=1, \ldots, 4}$ with Neumann boundary conditions and charges $(-1,0,0,0)$, $(1,0,1,-2)$, $(0,1,0,0)$, and $(0,-1,-1,4)$, and
$\Phi_{l=5, \ldots, 7}$ with Dirichlet boundary conditions and charges $(0,0,-1,4)$, $(-1,-1,-1,4)$, and $(-1,1,0,0)$.
The anomaly polynomial is zero, and there is no UV Chern-Simons term.
The superpotential\footnote{For a proper UV description of $T[M_3]$, relevant superpotential couplings should be specified, but it is difficult to determine them from the half-index from which the theory is engineered.
In this case, there are seven 3d chiral multiplets, and the superpotential \eqref{spot} preserves the global symmetries whose fugacities appear in the half-index.
The additional superpotential couplings needed to break extraneous global symmetries of the theory are not included in \eqref{spot}, and the corresponding fugacities are turned off by hand in the half-indices and other partition functions discussed in this work.
It is expected that there are theories with all necessary symmetry-breaking operators in their superpotentials, which would flow in the IR to the theories discussed in this work.
With the fugacities for extraneous global symmetries turned off, the protected quantities of the $T[M_3]$ discussed in this work, such as the SUSY parameter space and partition functions, are the same as those of such a theory.
} is given by
\begin{align}
W = \mu_1 \Phi_1 \Phi_2 \Phi_3 \Phi_4 + \mu_2 \Phi_1 \Phi_2 \Phi_5 + \mu_3 \Phi_2 \Phi_3 \Phi_6 + \mu_4 \Phi_2 \Phi_4 \Phi_7	\,	.
\label{spot}
\end{align}

Also, when $a=q^N$ with $N \in \mathbb{Z}_{\geq 0}$, we obtain the contour integral expression for the $SU(N)$ homological block, and the field content and boundary conditions of the corresponding 3d $\mathcal{N}=2$ theory $T[S^3 \backslash 4_1, SU(N)]$ can also be read off in a similar way.
In particular, when $a=q^N$, the $U(1)_a$ symmetry is not available, and the specialization $a=q^N$ is incorporated into charges and (mixed) Chern-Simons couplings.
For example, in this case, charges of $\Phi_{l=2,4,5,6}$ under $U(1)_z \times U(1)_x \times U(1)_R$ become $(1,0,2N-2)$, $(0,-1,4-2N)$, $(0,0,4-2N)$, and $(-1,-1,4-2N)$, respectively.	\\

In addition, by taking the poles $z=a^{-1}q^{1-k} = q^{1-N-k}$ with $N \in \mathbb{Z}_{\geq 2}$, $k=0,1, \ldots, n-1$, from $(q^{-1}az;q)^{-1}_\infty$, the contour integral \eqref{int41a} gives
\begin{align}
P_{4_1}(n,a,q) = \sum_{k=0}^{n-1} q^{\frac{1}{2}k(k+1)} \frac{(q^{n-k};q)_k}{(q;q)_k} (q^{-n+2-N-k};q)_k (q^{N-1};q)_k	\,	,
\label{homfly41}
\end{align}
and this agrees with the colored HOMFLY-PT polynomial \eqref{homfly} for the figure-eight knot.	\\

We can also consider another integral expression
\begin{align}
(q;q)_\infty \int \frac{dz}{2 \pi i z} \frac{1}{(z^{-1};q)_\infty} \frac{(q^{-1}a;q)_\infty}{(q^{-1}az;q)_\infty}
\frac{(qx^{-1};q)_\infty}{(q^{2}a^{-1}x^{-1}z^{-1};q)_\infty}
\frac{(q^{-1} ax;q)_\infty}{(z^{-1}x;q)_\infty}
\frac{\theta(z;q)}{\theta(1;q)} 
\frac{\theta(q^{-1}az;q)}{\theta(q^{-1}a;q)}	\,	.
\label{int41a-2}
\end{align}
As in the case of \eqref{int41a} with \eqref{integ41a}, the integral \eqref{int41a-2} also produces the homological block \eqref{hb41n}, or its $a$-deformed version \eqref{hb41a}, and the colored HOMFLY-PT polynomial \eqref{homfly41} by choosing the same sets of poles.
Compared to the half-index \eqref{int41a}, four 3d chiral multiplets, $\Phi_{l=3,4,6,7}$, whose charges are $(0,1,0,0)$, $(0,-1,-1,4)$, $(-1,-1,-1,4)$, and $(-1,1,0,0)$ switch their boundary conditions $N \leftrightarrow D$.
In addition, 2d $\mathcal{N}=(0,2)$ chiral and Fermi multiplets are introduced: 
chiral multiplets with charges $(0,0,0,1)$ and $(0,0,1,-1)$, and 
Fermi multiplets with charges $(1,0,1,-2)$ and $(1,0,0,0)$.
The superpotential is the same as in the case of \eqref{integ41a}.
When $a=q^N$, $N \in \mathbb{Z}_{\geq 2}$, a theory $T[S^3 \backslash 4_1, SL(N,\mathbb{C})]$ can also be obtained in a similar way.

We note that other choices of theta functions, \textit{i.e.} of 2d boundary degrees of freedom, can also be made in \eqref{int41a-2}, provided that they give the same homological block when $a=q^N$.
For generic $a$, other such choices of theta functions give homological blocks that agree with \eqref{hb41a} up to theta function ambiguity.\footnote{
Thus, when we express the $a$-deformed homological block in terms of theta functions, there is a theta function ambiguity. 
In general, if we want to avoid such an ambiguity, we may take the expression with $a=q^N$, which doesn't have this ambiguity, rewrite it so that all dependence on $N$ appears in exponents, and replace $N$ by $\frac{\log a}{\log q}$.
For the figure-eight knot, this gives \eqref{hb41a}.
For the trefoil knots, we refer to \eqref{hb31l-2} and \eqref{hb31r-2}.}
In general, such an ambiguity can arise in the half-index realization of the inverted Habiro series.	\\

In addition to the poles from $(z^{-1};q)_\infty^{-1}$, there are other sets of poles, which are $z=xq^k$, $k=0,1, \ldots$, from $(z^{-1}x;q)_\infty^{-1}$, and $z=q^{2}a^{-1} x^{-1} q^k$, $k=0,1, \ldots$, from $(q^{2}a^{-1}x^{-1}z^{-1};q)_\infty^{-1}$.
By taking the poles $z=xq^k$, $k=0,1, \ldots$, we obtain
\begin{align}
\frac{\theta(x;q)}{\theta(1;q)}
\frac{\theta(q^{-1}a x;q)}{\theta(q^{-1}a;q)}
\frac{(q^{-1}a;q)_\infty (qx^{-1};q)_\infty }{(x^{-1};q)_\infty (q^2 a^{-1} x^{-2};q)_\infty} 
\sum_{k=0}^{\infty}\frac{(-1)^k x^k q^{\frac{1}{2}k(k+1)} (q^{-1} a x;q)_k}{(q;q)_k  (qx;q)_k  (q^{-1}ax^2;q)_k} \,	.
\label{41nab1}
\end{align}
The contour integral with the poles $z=q^{2}a^{-1} x^{-1} q^k$, $k=0,1, \ldots$, is given by \eqref{41nab1} with $x$ replaced by $q^{2}a^{-1}x^{-1}$,
\begin{align}
\frac{\theta(q^2 a^{-1} x^{-1};q)}{\theta(1;q)}
\frac{\theta(q x^{-1};q)}{\theta(q^{-1}a;q)}
\frac{(q^{-1}a;q)_\infty (q^{-1} a x;q)_\infty }{(q^{-2} a x;q)_\infty (q^{-2} a x^2;q)_\infty} 
\sum_{k=0}^{\infty}\frac{(-1)^k a^{-k} x^{-k} q^{\frac{1}{2}k(k+5)} (q x^{-1};q)_k}{(q;q)_k  (q^3 a^{-1} x^{-1};q)_k  (q^{3} a^{-1} x^{-2};q)_k} \,	.
\label{41nab2}
\end{align}
The positive expansions of \eqref{41nab1} and \eqref{41nab2} agree with those obtained in \cite{EGGKPSS}.
Both are annihilated the same quantum $\hat{A}$ and $\hat{B}$-polynomials \eqref{qapoly41} and \eqref{qbpoly41}.
Also, for both lexicographic orders $a\succ x$ and $x\succ a$, the Gr\"{o}bner bases obtained for the half-indices \eqref{41nab1} and \eqref{41nab2} are both the same as those obtained above for the homological block $F_{4_1}$ in the discussion of the $\widehat{AB}_{4_1}$-ideal.

For $a=q^N$ with $N \in \mathbb{Z}_{\geq 2}$, the operator $\hat{A}_{4_1}^{\text{nab}}(N;\hat{x},\hat{y},q)$, which corresponds to the non-abelian branch, annihilates \eqref{41nab1} and \eqref{41nab2}.
For example, $\hat{A}_{4_1}^{\text{nab}}(N;\hat{x},\hat{y},q)$ for $N=3$ is given by \eqref{qapolynab3}.


\subsubsection*{Brane configurations for finite and large $N$}

The 3d $\mathcal{N}=2$ theories corresponding to the homological blocks for finite and large $N$ are realized in an M-theory configuration.
The M-theory configuration that realizes the 3d-3d correspondence for a knot complement at finite $N$ is given by
\begin{align}
\begin{tabular}{r c c c c c c}
\text{space-time}					&	&	$S^1$ 	&$\times$ 	&$TN$ 	&$\times$ 	&$T^* M_3$	\\
$N$ \text{M5 branes}				&	&	$S^1$ 	&$\times$ 	&$D^2$ 	&$\times$ 	&$M_3$	\\
$N^{\prime}$ \text{M5$^{\prime}$ branes}	&	&	$S^1$ 	&$\times$ 	&$D^2$ 	&$\times$ 	&$L_K$
\end{tabular}
\label{mconfn}
\end{align}
where $TN$ is a Taub-NUT space, $D^2$ is a disc in $TN$, and $T^* M_3$ is a cotangent bundle of $M_3$ \cite{Ooguri-Vafa, Witten-M5knots}.
We take $M_3=S^3$ in \eqref{mconfn}.
$L_K$ is a conormal bundle of a knot $K$ whose intersection with $M_3$ is $M_3 \cap L_K = K$ and it is a Lagrangian submanifold in $T^* M_3$.
Integrating out degrees of freedom along a knot $K$ effectively removes the knot in $S^3$, leading to the knot complement $S^3 \backslash K$, with boundary conditions on $T^2 = \partial (S^3 \backslash K)$.

The homological block for $G=SU(N)$ is given by the analytically continued Chern-Simons partition function on $S^3 \backslash K$ and corresponds to the half-index of a 3d $\mathcal{N}=2$ theory $T[M_3]$ on $D^2 \times_q S^1$.
In the partition functions, there are at most $N^{\prime}$ additional parameters, $x_j \in \mathbb{C}^*$, $j =1, \ldots, N^{\prime}$, arising from the holonomies of the $GL(N', \mathbb{C})$ flat connections.
These parameters describe the boundary conditions on $T^2 = \partial (S^3 \backslash K)$ for the $G_{\mathbb{C}}=SL(N,\mathbb{C})$ flat connections after decoupling the $GL(1,\mathbb{C})$ sector from $GL(N,\mathbb{C})$. 
When $N=2$, the partition functions for knot complements in the 3d-3d correspondence \cite{Dimofte-Gaiotto-Gukov, Beem-Dimofte-Pasquetti} arise from the configuration with two M5$^{\prime}$-branes on $L_K$, and the Weyl symmetry under $x \leftrightarrow x^{-1}$, which exchanges $x_1$ and $x_2$, is manifest.
For $N \geq 2$ and totally symmetric representations, the homological block would also arise from the configuration with two M5$^{\prime}$-branes on $L_K$, and the Weyl symmetry under $x \leftrightarrow q^{2-N} x^{-1}$ is manifest.	\\

After the large-$N$ limit, the M-theory configuration \eqref{mconfn} goes through the geometric transition and becomes
\begin{align}
\begin{tabular}{r c c c c c c}
\text{space-time}					&	&	$S^1$ 	&$\times$ 	&$TN$ 	&$\times$ 	&$X$	\\
$N^{\prime}$ \text{M5$^{\prime}$ branes}	&	&	$S^1$ 	&$\times$ 	&$D^2$ 	&$\times$ 	&$L_K$
\end{tabular}
\label{mconfa}
\end{align}
where $X$ is a resolved conifold $X := \mathcal{O}(-1) \oplus \mathcal{O}(-1) \rightarrow \mathbf{P}^1$.
We use the same notation $L_K$ for the conormal bundle of $K$, which is a special Lagrangian submanifold in $X$.

The 3d-3d correspondence in this setup relates the Chern-Simons theory on $L_K \subset X$ to the 3d $\mathcal{N}=2$ theory $T[L_K]$ on $D^2 \times_q S^1$. 
There is an additional parameter $a = q^{N}$ in this configuration, which is given by the exponential of the complexified K\"ahler parameter.
This parameter is understood as a fugacity of the $U(1)_a$ global symmetry associated with the internal 2-cycle in the resolved conifold $X$.

Also, before the transition, it is possible to connect $N'$ M5$'$-branes to $N'$ branes of $N$ M5-branes \cite{Aganagic-Vafa-Q, EGGKPS, EGGKPSS}.
This gives the $N'$ M5-branes supported on $M_K:=S^3 \backslash K$.
Assuming $N' \sim \mathcal{O}(1)$ as $N \rightarrow \infty$, the large-$N$ transition leads to the configuration \eqref{mconfa} with $L_K$ replaced by $M_K \subset X$, where we also use $M_K$ by abuse of notation, and the K\"ahler parameter is given by $a' = q^{N-N'}$.
The $a$-deformed homological block $F_K(x,a,q)$ with $a=q^2 a'$ would arise from the open topological string partition function on $X$ with branes on $M_K$, or equivalently the partition function of Chern-Simons theory on $M_K \subset X$.
The 3d-3d correspondence relates this with the half-index of $T[M_K]$ on $D^2 \times_q S^1$.\footnote{In general, the shift $a=q^2 a'$ affects the $R$-charge assignments of the fields and the (mixed) Chern-Simons couplings in the 3d $\mathcal{N}=2$ theories. 
It also affects the $R$-charges of monopole operators charged under $U(1)_a$, but the gauge-invariant monopole operators in the theories discussed in section \ref{ssec:31} are neutral under $U(1)_a$.}


\subsection{Homological block from a colored HOMFLY-PT polynomial}
\label{ssec:hbhp}

For general knots, closed-form expressions for the inverted Habiro coefficients in \eqref{hbn1} for general $N$ are not available yet. 
For the trefoil knots with $G=SU(N)$, for example, explicit expressions for these coefficients don't seem to be available in the literature.

For the figure-eight knot, the homological block $F_K$ and the colored HOMFLY-PT polynomial $P_K$ were obtained by taking the sets of poles $z=q^k$, $k=0,1, \ldots$, and  $z=q^{1-N-k}$, $k=0,1, \ldots, n-1$, respectively, from the integral \eqref{int41a} or \eqref{int41a-2}.
Therefore, by analyzing the homological block and the colored HOMFLY-PT polynomial obtained from the contour integral, we can obtain the homological block in the form of an inverted Habiro series from the expression for the colored HOMFLY-PT polynomial as a Habiro series.	\\

For simplicity, we take the form of the integral as in \eqref{int41a} with $a=q^N$ and $N \in \mathbb{Z}_{\geq 2}$, but with an extra factor $\Theta_z$,
\begin{align}
(q;q)_\infty \int \frac{dz}{2 \pi i z} \frac{1}{(z^{-1};q)_\infty} \frac{(q^{N-1};q)_\infty}{(q^{N-1}z;q)_\infty} \frac{(q^{N-1}xz;q)_\infty}{(x;q)_\infty} \frac{(q x^{-1} z;q)_\infty}{(q^{2-N} x^{-1} ;q)_\infty} \, \Theta_z
\label{int41-gen}
\end{align}
where $\Theta_z$ is a certain function that depends on $z$, $q$, $q^N$, and possibly $x$, such that when $z=q^k$, $\Theta_{z=q^k}$ doesn't have $x$-dependence.
Then, by taking the poles $z=q^k$, $k=0,1, \ldots$, the integral \eqref{int41-gen} becomes the infinite sum
\begin{align}
\sum_{k=0}^{\infty} \Theta_{z=q^k} \frac{(-1)^k q^{\frac{1}{2}k(k+1)}}{(x;q)_{k+N-1} (x^{-1} q^{2-N};q)_{k+N-1}} 
\frac{(q^{k+1};q)_\infty}{(q;q)_\infty}
\frac{(q^{N-1};q)_\infty}{(q^{k+N-1};q)_\infty}	\,	.
\label{cand}
\end{align}
We want $\Theta_{z=q^k}$ to give the correct inverted Habiro coefficients $\alpha_{-k-1}(K;N,q)$.
Then the sum \eqref{cand} gives the homological block \eqref{hbn1}.

By taking the poles $z=q^{1-N-k}$, $k=0, 1, \ldots, n-1$, in \eqref{int41-gen}, we obtain 
\begin{align}
\sum_{k=0}^{n-1} \Theta_{z=q^{1-N-k}} \, q^{\frac{1}{2}k(k+1)} \frac{(q^{n-k};q)_k}{(q;q)_k} (q^{-n} q^{2-N-k};q)_k (q^{N-1};q)_k	\,	
\label{hsfromint}
\end{align}
where $x=q^n$ is taken.
Meanwhile, the colored HOMFLY-PT polynomial \eqref{homfly} can be expressed as
\begin{align}
P_K(n,a,q) = \sum_{k=0}^{n-1} \mathfrak{a}_k(K;a,q) \, q^{\frac{1}{2}k(k+1)} \frac{(q^{n-k};q)_k}{(q;q)_k} (q^{-n} q^{2-N-k};q)_k (q^{N-1};q)_k	\,	.
\label{homfly2}
\end{align}
Therefore, \eqref{hsfromint} is identified with the colored HOMFLY-PT polynomial \eqref{homfly2}, if 
\begin{align}
\mathfrak{a}_k(K;a,q) \big|_{a=q^N} = \Theta_{z=q^{1-N-k}}	\,	.
\label{hcoeff}
\end{align}
From \eqref{hcoeff}, we can obtain $\Theta_z$.\footnote{It is possible that the inverted Habiro coefficients are given by the sums over certain discrete variables or contains $q$-Pochhammer symbols. Then $\Theta_z$ is expressed as contour integrals over the corresponding continuous variables or contains the corresponding $q$-Pochhammer symbols with appropriate continuous variables.}
Then, by evaluating $\Theta_{z=q^k}$, we would obtain the correct inverted Habiro coefficients, and accordingly the homological block in the form of an inverted Habiro series.	\\

Both the homological block $F_K$ and the colored HOMFLY-PT polynomial $P_K$ are expected to be annihilated by the same $q$-difference operators, up to additional annihilators that are associated with formal critical points at infinity.
We have checked these for the examples discussed in this work.
The method described above also applies to expressions with additional poles, for example, to the case in which $(q^{2-N}x^{-1}z^{-1};q)_\infty^{-1} (z^{-1}x;q)_\infty^{-1}$ appears in the integral as in \eqref{int41a-2}, \eqref{int31la-2}, and \eqref{int31ra-2}, and to the case in which $(z^{-1}x;q)_\infty^{-1}$ appears in the integral as in \eqref{int31l} and \eqref{int31r}.
Other than the $\Theta_z$ factor, their integrands differ by rational functions of theta functions, which can be absorbed into $\Theta_z$ in \eqref{int41-gen}.


\subsubsection*{Homological blocks on $|q|<1$ and $|q|>1$}

Extending the homological block from $|q|<1$ to $|q|>1$, or taking $q \rightarrow q^{-1}$, is subtle \cite{CCFGH}.
For example, at the level of series expansions, this operation amounts to replacing $q$ by $q^{-1}$ in a $q$-hypergeometric series expression for the homological block for $|q|<1$, when such an expression is available, and then expanding the resulting expression as a $q$-series. 
The resulting series would give the homological block for $|q|>1$.
However, such an extension is known to be ambiguous \cite{CCFGH}.

For the case of a knot $K$, the operation $q \rightarrow q^{-1}$ would give the homological block of a mirror knot $m(K)$.
Though it is subtle at the level of homological blocks, the operation $q \rightarrow q^{-1}$ can be performed on the colored HOMFLY-PT polynomial where $a=q^N$, and gives that of the mirror knot $m(K)$.
Therefore, since the homological block of a knot $K$ can be obtained from the colored HOMFLY-PT polynomial for $K$ as described above, we can also obtain the homological block for $m(K)$ from the colored HOMFLY-PT polynomial for $m(K)$.
By doing so, we obtain the extension of the homological block for $K$ to $|q|>1$.

We see that this is consistent with the case of the figure-eight knot, which is amphichiral, as well as with the cases of the left- and right-handed trefoil knots discussed below.
Also, if the homological block for a knot $K$ gives rise to a quantum modular form of a certain type, then, by analogy with the relation between false theta functions and mock modular forms appearing in the case of certain Seifert manifolds, the method described above would provide the corresponding quantum modular form on the $|q|>1$ side.

Given the homological block corresponding to the half-index of $T[M_3]$, extension of it to $|q|>1$ corresponds to the half-index of $T[-M_3]$.
Having both the homological blocks for a knot and its mirror knot, we can obtain both $T[S^3 \backslash K]$ and $T[S^3 \backslash m(K)]$ where $S^3 \backslash m(K)$ is the orientation reversal of $S^3 \backslash K$.


\subsection{Trefoil knots}
\label{ssec:31}

In this section, we discuss the case of the trefoil knots.
Since the homological blocks for the left- and the right-handed trefoil knots exhibit some differences, we discuss them separately.


\subsubsection*{Left-handed trefoil knot}

For the left-handed trefoil, the Habiro coefficients are given by $\mathfrak{a}_k(3_1^{l};a,q) = (-1)^k q^{\frac{1}{2}k(k-1)} a^k$.
By using the method described in section \ref{ssec:hbhp}, the inverted Habiro coefficients are obtained as
\begin{align}
\alpha_{-k-1}(3_1^l;N,q) = (-1)^{k+1} q^{\frac{1}{2}(-k-1)(-k+2)} (-q^{\frac{1}{2}})^{-3(N-2)} q^{-\frac{1}{2}(N-2)^2}	\,	,	\quad	k=0,1,\ldots	\,	.
\end{align}
Therefore, the homological block for $G=SU(N)$ is given by
\begin{align}
F_{3_1^l}^{SU(N)}(x,q) = 
-q^{-1} \sum_{k=0}^{\infty} \frac{q^{k^2} (-q^{\frac{1}{2}})^{-3(N-2)} q^{-\frac{1}{2}(N-2)^2} }{(x;q)_{k+N-1} (x^{-1} q^{2-N};q)_{k+N-1}} 
\frac{(q^{k+1};q)_\infty}{(q;q)_\infty}
\frac{(q^{N-1};q)_\infty}{(q^{k+N-1};q)_\infty}	\,	.
\label{hb31l}
\end{align}
When $N=2$, this reproduces the homological block for $G=SU(2)$ in the form of the inverted Habiro series.
When $N=1$, it is trivial, as expected.
The power series expansion of \eqref{hb31l} in $x$ agrees with the positive expansion obtained in \cite{EGGKPSS}.	\\

The homological block \eqref{hb31l} and the colored HOMFLY-PT polynomial \eqref{homfly2} for the left-handed trefoil knot can be obtained from the integral $(q;q)_\infty \int \frac{dz}{2 \pi i z} \Upsilon_{3_1^l}(z,x,a,q)$ with 
\begin{align}
\begin{split}
\Upsilon_{3_1^l}(z,x,a,q) &= \frac{1}{(z^{-1};q)_\infty} \frac{(q^{-1}a;q)_\infty}{(q^{-1}az;q)_\infty} \frac{(q^{-1}axz;q)_\infty}{(x;q)_\infty} 
\frac{(q^{-1}ax;q)_\infty}{(z^{-1} x;q)_\infty}	\\
&\hspace{30mm} \times \frac{\theta((-q^{\frac{1}{2}})^{-1} ax^{-3} z;q)}{\theta((-q^{\frac{1}{2}})x^{-3};q)}
\frac{\theta((-q^{\frac{1}{2}})^{-1} x^{-2};q)}{\theta((-q^{\frac{1}{2}})^{-3} ax^{-2} z;q)}
\label{int31l}
\end{split}
\end{align}
by taking the poles $z=q^{k}$, $k=0,1, \ldots$, with $x\in \mathbb{C}^*$, and $z=q^{1-k-N}$, $k = 0,1, \ldots, n-1$, with $x=q^n$, $n \in \mathbb{Z}_{\geq 1}$, respectively.
For generic $a$, the contour enclosing the poles $z=q^{k}$, $k=0,1, \ldots$, leads to
\begin{align}
\begin{split}
&F_{3_1^l}(x,a,q) \simeq \frac{\theta((-q^{\frac{1}{2}})^{-3} a x;q)}{\theta((-q^{\frac{1}{2}})x;q)}
\frac{\theta((-q^{\frac{1}{2}})^{-3} a x^{-3};q)}{\theta((-q^{\frac{1}{2}})x^{-3};q)}
\frac{\theta((-q^{\frac{1}{2}})^{-1}x^{-2};q)}{\theta((-q^{\frac{1}{2}})^{-5}ax^{-2};q)}	\\
&\hspace{30mm}\times
q^{-1} \frac{(q x^{-1};q)_\infty (q^{-1} a x;q)_\infty}{ (x;q)_\infty (q^2a^{-1}x^{-1};q)_\infty}
\sum_{k=0}^{\infty} 
\frac{q^{k^2} (q^{-1}a;q)_k }{(q;q)_k (q x^{-1};q)_k  (q^{-1} a x;q)_k }	\,	,
\end{split}
\label{hb31la}
\end{align}
which recovers \eqref{hb31l} when $a=q^N$.
Here, we use $\simeq$ to denote that the homological block $F_{3_1^l}(x,a,q)$ is given by the half-index on the right-hand side up to theta function ambiguity.\footnote{If we want to avoid theta function ambiguity, the homological block $F_{3_1^l}(x,a,q)$ can also be expressed as 
\begin{align}
F_{3_1^l}(x,a,q) = -q^{-1} \sum_{k=0}^{\infty} q^{k^2} (-q^{\frac{1}{2}})^{-3\big(\frac{\log a}{\log q}-2\big)} q^{-\frac{1}{2}\big(\frac{\log a}{\log q}-2\big)^2} 
\frac{(q^{k+1};q)_\infty}{(q;q)_\infty}
\frac{(q^{-1}a;q)_\infty}{(q^{k-1}a;q)_\infty}
\frac{(xaq^{k-1};q)_\infty}{(x;q)_\infty} \frac{(x^{-1} q^{k+1};q)_\infty}{(x^{-1} a^{-1}q^{2};q)_\infty}		\,	.
\label{hb31l-2}
\end{align}}

The 3d $\mathcal{N}=2$ theory whose half-index is \eqref{int31l} is given by the following field contents under $U(1)_z \times U(1)_x \times U(1)_a \times U(1)_R$ symmetries with boundary conditions: 
a $U(1)_z$ vector multiplet with Neumann boundary condition;
the seven 3d chiral multiplets, $\Phi_{l=1, \ldots, 7}$, where
$\Phi_{l=1, 2, 3, 7}$ with Neumann boundary conditions and charges $(-1,0,0,0)$, $(1,0,1,-2)$, $(0,1,0,0)$, and $(-1,1,0,0)$, and
$\Phi_{l=4, 5, 6}$ with Dirichlet boundary conditions and charges $(0,-1,-1,4)$, $(0,0,-1,4)$, and $(-1,-1,-1,4)$;
2d chiral multiplets with charges $(0,-3,0,2)$ and $(1,-2,1,2)$; and
2d Fermi multiplets with charges $(1,-3,1,-1)$ and $(0,-2,0,-1)$.
The anomaly is cancelled by introducing the UV Chern-Simons terms encoded in $\mathbf{f}_z^2 - \mathbf{f}_a^2 - 2\mathbf{f}_R(\mathbf{f}_z - \mathbf{f}_a)$, where $\mathbf{f}_*$ denotes the field strength of $U(1)_*$.	
The superpotential is given by \eqref{spot}.
There is a gauge-invariant anti-monopole operator with charges $(0,0,0,4)$.
For $a=q^N$, the corresponding 3d $\mathcal{N}=2$ theory $T[S^3 \backslash 3_1^l, SU(N)]$ can be obtained similarly.	\\

The effective twisted superpotential for the left-handed trefoil knot is given by
\begin{align}
\begin{split}
&\widetilde{\mathcal{W}}_{3_1^l}(z,x,a) = (\log z)^2 + \log(-1) \log a -\frac{1}{2} (\log a)^2 + \text{Li}_2(z x a) - \text{Li}_2(x)	\\
&\hspace{30mm} + \text{Li}_2(x^{-1} z) - \text{Li}_2(x^{-1} a^{-1}) + \text{Li}_2(z) - \text{Li}_2(z a) + \text{Li}_2(a)	\,	.
\end{split}
\end{align}
From the critical point equation \eqref{svac},
\begin{align}
1=\frac{z^2 (1- a z)}{(1-z) (1-x^{-1}z) (1-a x z)}	\,	,	\label{svac31l}
\end{align}
and the equations for the supersymmetric parameter space \eqref{spar}, we have 
\begin{align}
\begin{split}
&\hspace{-2mm} 0 = \widetilde{A}_{3_1^l}(x,y,a) = \big(x (a x-1) y +x-1\big)	\\
&\hspace{16mm} \times \big( (a x-1) y^2 + (a^3 x^4-a^2 x^3-2 (a-1) a x^2-a x+a ) y	-a^2 (x-1) x^3 \big)	\,	,
\end{split}	\label{apoly31ltw}	\\
\begin{split}
&\hspace{-2mm} 0 = \widetilde{B}_{3_1^l}(a,b,x) = \big( (a-1)(x a-1)b-1 \big)	\\
&\hspace{30mm} \times \big( (a-1) (x a - 1) b^2 - x (x^2 a^2 - (x+1)a + 2) b + x^2 \big)	\,	,
\end{split}	\label{bpoly31ltw}	
\end{align}
where the second factors on the right-hand sides of \eqref{apoly31ltw} and \eqref{bpoly31ltw} are the classical $a$-deformed $A$-polynomial and the $x$-deformed $B$-polynomial, respectively.
When taking $a=1$ in \eqref{apoly31ltw} and $x=1$ in \eqref{bpoly31ltw}, they become
\begin{align}
\widetilde{A}_{3_1^l}(x,y,1)	&= (x-1)^2 (x y+1) (y-1) (y+x^3)	\,	,			\label{poly31ltwa1}\\
\widetilde{B}_{3_1^l}(a,b,1)	&= (b-1) \big((a-1)^2 b-1\big)^2	\,	,		\label{poly31ltwx1}
\end{align}
respectively, where $y-1$ and $b-1$ correspond to the abelian branch, and $y+x^3$ and $(a-1)^2 b-1$ correspond to the non-abelian branch.
The factors $xy+1$ and $(a-1)^2 b-1$ of \eqref{poly31ltwa1} and \eqref{poly31ltwx1}, which are obtained from the first factors on the right-hand sides of \eqref{apoly31ltw} and \eqref{bpoly31ltw}, respectively, arise from the formal solution $z=\infty$ of the supersymmetric vacuum condition \eqref{svac31l}.	\\

The $a$-deformed homological block \eqref{hb31la} is annihilated by the quantum $\hat{A}$- and $\hat{B}$-polynomials
\begin{align}
\begin{split}
&\hat{A}_{3_1^l}(\hat{x},\hat{y},a,q) = 
q (a \hat{x}-1) (a \hat{x}^2-q) \hat{y}^2 -a^2 \hat{x}^3 q (\hat{x}-1) (a q \hat{x}^2-1)	\\
&\hspace{22.5mm}+ a (a \hat{x}^2-1) \left( a^2 q \hat{x}^4-a q^2 \hat{x}^3+ (q^2-a (1+q^2)+q^3 )\hat{x}^2 -q^2 \hat{x}+q \right) \hat{y}	\,	,
\end{split}	
\label{qapoly31l}	\\
&\hat{B}_{3_1^l}(\hat{a},\hat{b},x,q) =
q^2 (\hat{a}-1) (x\hat{a}-1) \hat{b}^2	
- x (x^2\hat{a}^2 -(1+x)q \hat{a} +q^2+q) \hat{b}	
+q x^2	\,	,
\label{qbpoly31l}
\end{align}
which agree with \cite{FGS-superA, EGGKPSS}.

The classical limits of the quantum $\hat{A}$-polynomial \eqref{qapoly31l} and $\hat{B}$-polynomial \eqref{qbpoly31l} are given by
\begin{align}
\begin{split}
A_{3_1^l}(x,y,a)/(a x^2-1) &= 
(a x-1) y^2 
+ (a^3 x^4-a^2 x^3-2 (a-1) a x^2-a x+a ) y	\\
&\hspace{75mm}-a^2 (x-1) x^3	\,	,
\end{split}
\label{apoly31l}
\\
B_{3_1^l}(a,b,x) &= (a-1) (xa-1) b^2 
- x(x^2a^2 -(x+1)a +2) b
+ x^2	\,	,
\label{bpoly31l}
\end{align}
which agree with the classical $a$-deformed $A$-polynomial and $x$-deformed $B$-polynomial in \eqref{apoly31ltw} and \eqref{bpoly31ltw}, respectively.

As discussed in \cite{Chung-ab}, the annihilator associated with the formal solution $z=\infty$ of the supersymmetric vacuum equation \eqref{svac31l} doesn't arise in the operator annihilating the homological block \eqref{hb31la}, since it is obtained from the contour enclosing the poles $z=q^k$, $k=0,1, \ldots$, and extending toward $0$ in the $z$-plane.
Also, as in the case of $G=SU(2)$ \cite{Chung-ab}, applying the quantum $\hat{A}$-polynomial \eqref{qapoly31l} to the expression obtained from the colored HOMFLY-PT polynomial \eqref{homfly2}, with the summation range extended to infinity and with generic $x=q^n$, where $n$ is analytically continued, leads to an inhomogeneous equation.
Here, this infinite sum with generic $x$ is obtained by choosing the contour that encloses the poles $z=a^{-1}q^{1-k}$, $k=0,1, \ldots$, and extends toward $\infty$ in the $z$-plane.
The inhomogeneous term is annihilated by $\hat{x} (a \hat{x}-1) (a \hat{x}^2-1) (a \hat{x}^2-q) \hat{y} + (q \hat{x}-1) (a q^2 \hat{x}^2-1) (a q^3 \hat{x}^2-1)$ whose classical limit agrees, up to an overall factor, with $xy+1$ in \eqref{poly31ltwa1}, which is associated with the formal solution $z=\infty$ in \eqref{svac31l}.
Also, when $x=q^n$ with $n \in \mathbb{Z}_{\geq1}$, the inhomogeneous term vanishes, and the additional operator doesn't arise in the annihilator of the HOMFLY-PT polynomial colored by the totally symmetric representation with $n-1$ boxes.	\\

For $N \in \mathbb{Z}_{\geq 2}$, applying the operator $\hat{A}_{3_1^l}^{\text{nab}}(N;\hat{x},\hat{y},q)$, which corresponds to the non-abelian branch, to \eqref{hb31l} leads to an inhomogeneous equation.
The inhomogeneous term is annihilated by another operator $\hat{A}_{3_1^l}^{\text{ab}}(N;\hat{x},\hat{y},q)$, which corresponds to the abelian branch.
For example, when $N=3$, 
\begin{align}
\hat{A}_{3_1^l}^{\text{nab}}(3;\hat{x},\hat{y},q)	&= (q^2 \hat{x} - 1) (q \hat{x}^2 - q \hat{x} + 1) \hat{y} + q^3 \hat{x}^3 (\hat{x}-1) (q^3 \hat{x}^2 - q^2 \hat{x} + 1)	\,	,	\label{qapoly31l3nab}	\\
\hat{A}_{3_1^l}^{\text{ab}}(3;\hat{x},\hat{y},q)	&= (q^2 \hat{x}^2 - 1) \hat{y} + q^2 (1 - q^4 \hat{x}^2)	\,	,	\label{qapoly31l3ab}
\end{align}
and their classical limits are
\begin{align}
A_{3_1^l}^{\text{nab}}(3;x,y,q)	= (x-1) (x^2-x+1) (y+x^3)	\,	,	\quad
A_{3_1^l}^{\text{ab}}(3;x,y,q)	= (x^2-1) (y-1)	\,	.
\end{align}
Indeed, they correspond to non-abelian and abelian branches, respectively. 
As in the case of the figure-eight knot, by checking several values of $N$, we see that $\hat{A}_{3_1^l}^{\text{ab}}(N;\hat{x},\hat{y},q)\hat{A}_{3_1^l}^{\text{nab}}(N;\hat{x},\hat{y},q)$ agrees with $\hat{A}_{3_1^l}(\hat{x},\hat{y},a,q)$ at $a=q^N$ up to an overall factor, and this agreement should hold for general $N$. 	\\

We also obtain the $\widehat{AB}_{3_1^l}$-ideal (or its subideal) generated by the Gr\"obner basis $\{ \hat{D}_{3_1^l} , \hat{A}_{3_1^l}\}$ with respect to the lexicographic order $\hat{a} \succ \hat{x}$, where
\begin{align}
\hat{D}_{3_1^l}(\hat{x}, \hat{a}, \hat{y}, \hat{b},q) = (\hat{a}-q) (\hat{a} \hat{x}^2-q) \hat{a} + q \hat{x} \hat{y} - q \hat{x}	\,	.
\end{align}
For the lexicographic order $\hat{x} \succ \hat{a}$, the Gr\"obner basis is given by $\{ \hat{D}_{3_1^l} , \hat{B}_{3_1^l}\}$, and this agrees with that obtained in \cite{EGGKPSS}.	\\

When taking the poles $z=x q^k$, $k=0,1, \ldots$, the sum of the residues gives
\begin{align}
\begin{split}
&\mathcal{B}_{3_1^l}(x,a,q) = \frac{\theta((-q^{\frac{1}{2}})^{-1}a x^{-2};q)}{\theta((-q^{\frac{1}{2}})x^{-3};q)}
\frac{\theta((-q^{\frac{1}{2}})^{-1} x^{-2};q)}{\theta((-q^{\frac{1}{2}})^{-3} a x^{-1};q)}	
\frac{(q^{-1}a;q)_\infty (q^{-1} ax^2;q)_\infty}{(x;q)_\infty (x^{-1};q)_\infty }	\\
&\hspace{80mm}\times \sum_{k=0}^{\infty}
\frac{x^{2k} q^{k^2} (q^{-1} ax;q)_k}{(q;q)_k (qx;q)_k (q^{-1} ax^2;q)_k }  	\,	,
\end{split}
\label{31lnab}
\end{align}
and it is annihilated by the quantum $\hat{A}$ and $\hat{B}$-polynomials \eqref{qapoly31l} and \eqref{qbpoly31l}. 
Also, the Gr\"{o}bner bases obtained for \eqref{31lnab} are the same as those obtained above for the homological block $F_{3_1^l}$ in the discussion of the $\widehat{AB}_{3_1^l}$-ideal.
When $a=q^N$ with $N \in \mathbb{Z}_{\geq 2}$, \eqref{31lnab} is annihilated by the operator $\hat{A}_{3_1^l}^{\text{nab}}(N;\hat{x}, \hat{y},q)$ for the non-abelian branch.
For example, when $N=3$, this operator is given by \eqref{qapoly31l3nab}. 	\\

Applying the Weyl action $x \rightarrow q^2 a^{-1}x^{-1}$ to $\mathcal{B}_{3_1^l}(x,a,q)$ gives the half-index $\mathcal{B}_{3_1^l}(q^2 a^{-1}x^{-1},a,q)$, and both half-indices are annihilated by the same quantum $\hat{A}$- and $\hat{B}$-polynomials. 
When $a=q^N$, by comparing their $q$-series expansions for several values of $N$, we see that $\mathcal{B}_{3_1^l}(x,a,q)$ agrees with $\mathcal{B}_{3_1^l}(q^2 a^{-1}x^{-1},a,q)$.
For generic $a$, they agree up to theta function ambiguities.
The same holds for the case of the homological block \eqref{hb31la}.
When $a=q^N$, it is manifestly Weyl symmetric, but for generic $a$, it is Weyl symmetric up to theta function ambiguity.

In this case,\footnote{Though in this case $\mathcal{B}_{3_1^l}(x,a,q)$ and its Weyl partner agree up to theta function ambiguity and would agree for $a=q^N$, we discuss the relevant aspects in a more general setting.
For example, in the case of the right-handed trefoil knot discussed below, $\mathcal{B}_{3_1^r}(x,a,q)$ and its Weyl partner don't agree even when $a=q^N$.

Also, the same discussion can be applied to the integral with the integrand obtained from \eqref{int31l} by applying the Weyl action $x \rightarrow q^2 a^{-1}x^{-1}$.
Taking the residues at the poles $z= q^2 a^{-1}x^{-1} q^k$, $k=0, 1, \ldots$, in this integral gives $\mathcal{B}_{3_1^l}(q^2 a^{-1}x^{-1},a,q)$.
} 
the critical point equation \eqref{svac31l} has two solutions other than the formal solution $z =\infty$.
One would correspond to the contour that gives the homological block, while the other would correspond to the contour that gives \eqref{31lnab}. 
Since the system has Weyl symmetry, one may expect that the Weyl partner of \eqref{31lnab} also arises from the integral with the integrand \eqref{int31l}.\footnote{The Weyl partner also contributes to the partition functions in \eqref{indfact31l}.} 
Considering that, in general, the contour giving a half-index need not be a contour enclosing a set of poles \cite{Beem-Dimofte-Pasquetti}, though we don't specify the contour that gives the Weyl partner in the integral \eqref{int31l}, we expect that the Weyl partner of \eqref{31lnab} would also be obtained from an appropriately chosen contour.\footnote{Given that such a contour exists, a contour passing through one of the critical points would be expected to be either the contour giving \eqref{31lnab} or the contour giving its Weyl partner, depending on the values of the parameters $x$ and $a$.}

With this in mind, we can also consider another integral expression,
\begin{align}
\begin{split}
&\hspace{-1mm}(q;q)_\infty \int \frac{dz}{2 \pi i z} \frac{1}{(z^{-1};q)_\infty} \frac{(q^{-1}a;q)_\infty}{(q^{-1}az;q)_\infty}
\frac{(qx^{-1};q)_\infty}{(q^{2}a^{-1}x^{-1}z^{-1};q)_\infty}
\frac{(q^{-1} ax;q)_\infty}{(z^{-1}x;q)_\infty}	\\
&\hspace{9mm} \times 
\frac{\theta((-q^{\frac{1}{2}})x^2 z;q)}{\theta((-q^{\frac{1}{2}}) x^4 z;q)} 
\frac{\theta((-q^{\frac{1}{2}})^{-1}a x^2 z;q)}{\theta((-q^{\frac{1}{2}})^{-3} a x^2;q)} 
\frac{\theta((-q^{\frac{1}{2}})^{-3} a x^{-4};q)}{\theta((-q^{\frac{1}{2}})^{-3} a x^{-2};q)} 
\frac{\theta((-q^{\frac{1}{2}})^{-3} a x^{2};q)}{\theta((-q^{\frac{1}{2}}) x^{2};q)}	\,	,	
\end{split}
\label{int31la-2}
\end{align}
which is similar to the integral \eqref{int41a-2} in the case of the figure-eight knot.
The boundary conditions for the 3d chiral multiplets, $\Phi_{l=1, \ldots, 7}$, are the same as in the case of \eqref{int41a-2}, while \eqref{int31la-2} involves different 2d boundary multiplets.
The integral \eqref{int31la-2} gives the same half-indices as those obtained from \eqref{int31l}, including the homological block, up to theta function ambiguity for generic $a$.
In this case, in addition to the poles $z=q^k$, $k=0,1, \ldots$, associated with the homological block, there are two sets of poles, $z=xq^k$ and $z=q^2 a^{-1}x^{-1}q^k$, $k=0,1, \ldots$, whose residues give the Weyl pair $\mathcal{B}_{3_1^l}(x,a,q)$ and $\mathcal{B}_{3_1^l}(q^2 a^{-1}x^{-1},a,q)$, respectively, up to theta function ambiguities.
Thus, the Weyl partner of $\mathcal{B}_{3_1^l}(x,a,q)$ can be manifestly obtained from the residues in this case.\footnote{Also, a set of poles that a contour would enclose depends on the value of the critical point $z_*=z(x,a)$ through which the contour passes.}


\subsubsection*{Right-handed trefoil knot}

For the right-handed trefoil knot, the Habiro coefficients in \eqref{homfly2} are given by $\mathfrak{a}_k(3_1^r;a,q) = (-1)^k q^{-\frac{1}{2}k(k-1)} a^{-k}$.
As in the case of the left-handed trefoil knot, we obtain the inverted Habiro coefficients for the right-handed trefoil knot,
\begin{align}
\alpha_{-k-1}(3_1^r;N,q) = (-1)^{k+1} q^{-\frac{1}{2}(-k-1)(-k+2)} (-q^{\frac{1}{2}})^{3(N-2)} q^{\frac{1}{2}(N-2)^2}	\,	,
\label{ihc31r}
\end{align}
where we note that $\alpha_{-k-1}(3_1^r;N,q) = \alpha_{-k-1}(3_1^l;N,q^{-1})$.
Then the homological block for the right-handed trefoil knot is given by
\begin{align}
\hspace{-1mm}F_{3_1^r}^{SU(N)}(x,q) = -q \sum_{k=0}^{\infty} \frac{q^{k} (-q^{\frac{1}{2}})^{3(N-2)} q^{\frac{1}{2}(N-2)^2}}{(x;q)_{k+N-1} (x^{-1} q^{-(N-2)};q)_{k+N-1}} 
\frac{(q^{k+1};q)_\infty}{(q;q)_\infty}
\frac{(q^{N-1};q)_\infty}{(q^{k+N-1};q)_\infty}	\,	.
\label{hb31r}
\end{align}
When $N=2$, this agrees with the $G=SU(2)$ homological block in the form of the inverted Habiro series \cite{Park-inverted}.
For $N=1$, it is trivial, as expected.
The power series expansion of \eqref{hb31r} in $x$ also agrees with the positive expansion obtained in \cite{EGGKPSS}. 	\\

The homological block \eqref{hb31r} and the colored HOMFLY-PT polynomial \eqref{homfly2} with the Habiro coefficients $\mathfrak{a}_k(3_1^r;a,q)$ are obtained from the contour integral
\begin{align}
\begin{split}
&(q;q)_\infty \int \frac{dz}{2 \pi i z} \frac{1}{(z^{-1};q)_\infty} \frac{(q^{-1}a;q)_\infty}{(q^{-1}az;q)_\infty} \frac{(q^{-1}axz;q)_\infty}{(x;q)_\infty} 
\frac{(q^{-1}ax;q)_\infty}{(z^{-1} x;q)_\infty}	\\
&\hspace{30mm}\times 
\frac{\theta(q^{3} a^{-1} x z^{-1};q)}{\theta(q^2x;q)}
\frac{\theta(q^{-1} a z;q)}{\theta(1;q)}
\frac{\theta(q^{2} a^{-1} x^2;q)}{\theta(q^{3} a^{-2} x^2 z^{-1};q)}
\frac{\theta(q^{-1}x^2 z^{-1};q)}{\theta(q^{-2} a x^2;q)}
\label{int31r}
\end{split}
\end{align}
by taking the poles $z=q^{k}$, $k=0,1, \ldots$, with $x \in \mathbb{C}^*$, and $z=q^{1-k}a^{-1}$, $k = 0,1, \ldots, n-1$, with $x=q^n$ and $n \in \mathbb{Z}_{\geq 1}$, respectively. 
For generic $a$, the contour enclosing the poles $z=q^k$, $k=0,1, \ldots$, gives
\begin{align}
\begin{split}
&\hspace{-2mm}F_{3_1^r}(x,a,q) \simeq -x^{-1}
\frac{\theta(q^{3} a^{-1} x;q)}{\theta(q^2 x;q)}
\frac{\theta(q^{-1} a;q)}{\theta(1;q)}
\frac{\theta(q^{2} a^{-1} x^2;q)}{\theta(q^{3} a^{-2} x^2;q)}
\frac{\theta(q^{-1}x^2;q)}{\theta(q^{-2} a x^2;q)}	\\
&\hspace{12mm}\times\frac{\theta((-q^{\frac{1}{2}})^{-3} a x;q)}{\theta((-q^{\frac{1}{2}})x;q)}	
\frac{(qx^{-1};q)_\infty (q^{-1}ax;q)_\infty}{(x;q)_\infty (q^2 a^{-1}x^{-1};q)_\infty}
\sum_{k=0}^{\infty} \frac{q^k (q^{-1}a;q)_k}{(q;q)_k (qx^{-1};q)_k (q^{-1}ax;q)_k} 	\,	,
\end{split}	\label{hb31ra}
\end{align}
which recovers \eqref{hb31r} when $a=q^N$.\footnote{Without theta functions, the homological block $F_{3_1^r}(x,a,q)$ can also be expressed as 
\begin{align}
F_{3_1^r}(x,a,q) = -q \sum_{k=0}^{\infty} q^{k} (-q^{\frac{1}{2}})^{3\big(\frac{\log a}{\log q}-2\big)} q^{\frac{1}{2}\big(\frac{\log a}{\log q}-2\big)^2}
\frac{(q^{k+1};q)_\infty}{(q;q)_\infty}
\frac{(q^{-1}a;q)_\infty}{(q^{k-1}a;q)_\infty}
\frac{(xaq^{k-1};q)_\infty}{(x;q)_\infty} \frac{(x^{-1} q^{k+1};q)_\infty}{(x^{-1} a^{-1}q^{2};q)_\infty}		\,	.
\label{hb31r-2}
\end{align}}

The 3d $\mathcal{N}=2$ theory whose half-index is \eqref{int31r} is given by the following field contents under $U(1)_z \times U(1)_x \times U(1)_a \times U(1)_R$ symmetries with boundary conditions: 
a $U(1)_z$ vector multiplet with Neumann boundary condition;
the same seven 3d chiral multiplets as in the case of the left-handed trefoil knot;
2d chiral multiplets with charges $(0,1,0,5)$, $(0,0,0,1)$, $(-1,2,-2,7)$, and $(0,2,1,-3)$; and
2d Fermi multiplets with charges $(-1,1,-1,6)$, $(1,0,1,-2)$, $(0,2,-1,4)$, and $(-1,2,0,-2)$.
The UV Chern-Simons terms encoded in $-\mathbf{f}_z^2 + \mathbf{f}_a^2 + 2\mathbf{f}_R(\mathbf{f}_z - \mathbf{f}_a)$ are introduced to cancel anomaly.
There is a gauge-invariant monopole operator $V_+$ with charges $(0,0,0,2)$.
The superpotential is given by \eqref{spot}, and the monopole operator $V_+$ may also be included as a superpotential term.
For $a=q^N$, the corresponding 3d $\mathcal{N}=2$ theory $T[S^3 \backslash 3_1^r, SU(N)]$ can be obtained in a similar way.	\\

The effective twisted superpotential for the right-handed trefoil knot is given by
\begin{align}
\begin{split}
&\widetilde{\mathcal{W}}_{3_1^r}(z,x,a) = \log(-1) \log a + \frac{1}{2} (\log a)^2 + \text{Li}_2(z x a) - \text{Li}_2(x)	\\
&\hspace{25mm} + \text{Li}_2(x^{-1} z) - \text{Li}_2(x^{-1} a^{-1}) + \text{Li}_2(z) - \text{Li}_2(z a) + \text{Li}_2(a)	\,	.
\label{twspot31r}
\end{split}
\end{align}
From the supersymmetric vacuum equation \eqref{svac},
\begin{align}
1=\frac{(1 -a z )}{(1-z) (1-x^{-1}z) (1-a x z)}	\,	,	\label{svac31r}
\end{align}
and the SUSY parameter space \eqref{spar}, we have
\begin{align}
\begin{split}
&\hspace{-2mm}0 = \widetilde{A}_{3_1^r}(x,y,a) = \big((a x - 1) y + a x (x-1) \big)	\\
&\hspace{17mm}\times \big(a^2 x^3 (a x-1) y^2 - ( a^2 x^4-a^2 x^3+2 (a-1) a x^2-a x+1 ) y -a x+a \big)	\,	,	
\end{split}	\label{aapoly31rtw}	\\
\begin{split}
&\hspace{-2mm}0 = \widetilde{B}_{3_1^r}(a,b,x) = \big( (a-1)(a x-1) b - q^2 x a^2 \big)	\\
&\hspace{20mm}\times \big( x^2 (a-1)(a x-1) b^2 - \big(2 x^2 a^2 - x(x+1)a + 1 \big) b + x a^2 \big)	\,	.
\end{split}	\label{xbpoly31rtw}
\end{align}
The second factors on the right-hand sides of \eqref{aapoly31rtw} and \eqref{xbpoly31rtw} are the classical $a$-deformed $A$-polynomial and $x$-deformed $B$-polynomial for the right-handed trefoil knot, respectively.
Upon taking $a=1$ in \eqref{aapoly31rtw} and $x=1$ in \eqref{xbpoly31rtw}, we have
\begin{align}
\widetilde{A}_{3_1^r}(x,y,1) 	&= -(x-1)^2 (y+x) (y-1) (x^3 y+1)	\,	,	\label{apoly31rtw}	\\
 \widetilde{B}_{3_1^r}(a,b,1)	&= (b-1) \big( (a-1)^2 b-a^2 \big)^2	\,	,	\label{bpoly31rtw}
\end{align}
respectively, where $y-1$ and $b-1$ correspond to the abelian branch and $x^3 y+1$ and $(a-1)^2 b-a^2$ correspond to the non-abelian branch.
$y+x$ and $(a-1)^2 b-a^2$ in \eqref{apoly31rtw} and \eqref{bpoly31rtw}, which are obtained from the first factors on the right-hand sides of \eqref{aapoly31rtw} and \eqref{xbpoly31rtw}, respectively, arise from the formal solution $z=0$ of the critical point equation \eqref{svac31r}.	\\

The homological block \eqref{hb31r}, or \eqref{hb31ra}, is annihilated by
\begin{align}
\hat{A}^{'}_{3_1^r}(\hat{x},\hat{y},a,q) &= \hat{A}^{-\infty}_{3_1^r}(\hat{x},\hat{y},a,q) \hat{A}_{3_1^r}(\hat{x},\hat{y},a,q)	\,	,	\label{qapoly31rt}	\\
\hat{B}^{'}_{3_1^r}(\hat{a},\hat{b},x,q) &= \hat{B}^{-\infty}_{3_1^r}(\hat{a},\hat{b},x,q) \hat{B}_{3_1^r}(\hat{a},\hat{b},x,q)	\,	,	\label{qbpoly31rt}	
\end{align}
where\footnote{For reference, they can also be written as
\begin{align}
\begin{split}
&\hat{A}_{3_1^r}^{-\infty} \hat{A}_{3_1^r} = q (a q \hat{x}^2-1) \Big( 
-a^2 q^4 \hat{x}^3 (a \hat{x}-1) (a \hat{x}^2-1) (a q \hat{x}-1) (a \hat{x}^2-q) \hat{y}^3	\\
&\hspace{15mm}-q (a \hat{x}-1) (a \hat{x}^2-q) (a q^2 \hat{x}^2-1) 	\\
&\hspace{20mm}\times (a^3 q^3 \hat{x}^5-a^2 q^2 \hat{x}^4 (a+q)+a^2 (q^2+q-1) \hat{x}^3+a \hat{x}^2 (q^3-a (q+1)+1)+a \hat{x}-1) \hat{y}^2	\\
&\hspace{15mm}+a (a \hat{x}^2-1) (q \hat{x}-1)	(a q^3 \hat{x}^2-1) \\
&\hspace{20mm}\times (a^2 q^3 \hat{x}^5-a^2 q^2 \hat{x}^4+a q \hat{x}^3 (a (q+1)-q^3-1)-a (q^2+q-1) \hat{x}^2+q \hat{x} (a+q)-q) \hat{y}	\\
&\hspace{15mm}+a^2 q \hat{x} (\hat{x}-1) (q \hat{x}-1) (a q^2 \hat{x}^2-1) (a q^3 \hat{x}^2-1) \Big)	\,	,
\end{split}	\\
\begin{split}
&\hat{B}_{3_1^r}^{-\infty} \hat{B}_{3_1^r} = 
x^2 (\hat{a}-1) (q\hat{a} -1) (x\hat{a} -1) (q x\hat{a} -1) \hat{b}^3	\\
&\hspace{15mm} - q^2 (\hat{a}-1) (x\hat{a} -1) ( x^2 (q+x+1)\hat{a}^2 -  x(x+1) \hat{a}+1) \hat{b}^2	\\
&\hspace{15mm} + q^2 x \hat{a}^2 (((q+1) x^2+q x)\hat{a}^2  - q (x+1)^2 \hat{a} + q^2+q ) \hat{b}
-q^3 x^2 \hat{a}^4 	\,	.
\end{split}
\end{align}}
\begin{align}
\begin{split}
&\hspace{-4mm}\hat{A}_{3_1^r}(\hat{x},\hat{y},a,q) = a^2 q \hat{x}^3 (a \hat{x} - 1) (a \hat{x}^2 - q) \hat{y}^2 
-a q (\hat{x} - 1) (a q \hat{x}^2 - 1)	\\
&\hspace{14mm} - (a \hat{x}^2 - 1) \big(a^2 q^2 \hat{x}^4 + \hat{x}^2 \big( (q+1)a^2 - q (q^2+1)a  \big) - a^2 q \hat{x}^3 - a q \hat{x} + q^2 \big) \hat{y}	\,	,
\end{split}									\label{qapoly31r}	\\
\begin{split}
&\hspace{-4mm}\hat{A}^{-\infty}_{3_1^r}(\hat{x},\hat{y},a,q) = (a \hat{x} - 1) (a \hat{x}^2 - 1) (a \hat{x}^2 - q) \hat{y}  + a q \hat{x} (q \hat{x} - 1) (a q^2 \hat{x}^2 - 1) (a q^3 \hat{x}^2 - 1)	\,	,	\label{qapoly31ri}
\end{split}
\end{align}
and
\begin{align}
&\hat{B}_{3_1^r}(\hat{a},\hat{b},x,q) = 
(\hat{a} - 1) (x \hat{a} - 1) x^2 \hat{b}^2 
- \big( x^2 (q+1) \hat{a}^2  - q x (x+1) \hat{a} + q^2 \big) \hat{b}
+q x\hat{a}^2	\,	,							\label{qbpoly31r}	\\
&\hat{B}^{-\infty}_{3_1^r}(\hat{a},\hat{b},x,q) = (\hat{a} - 1) (x \hat{a} - 1) \hat{b} - q^2 x \hat{a}^2	\,	.	\label{qbpoly31ri}
\end{align}
The operator \eqref{qapoly31r} agrees with the quantum $\hat{A}$-polynomial in \cite{EGGKPSS}.

The classical limits of the operators \eqref{qapoly31rt} and \eqref{qbpoly31rt} for the $a$-deformed homological block are
\begin{align}
\begin{split}
&\hspace{-1mm} A_{3_1^r}(x,y,a) / (a x^2 - 1) 	\\
&\hspace{17mm} = a^2 x^3 (a x-1) y^2 - ( a^2 x^4-a^2 x^3+2 (a-1) a x^2-a x+1 ) y -a x+a	\,	,
\end{split}		\label{apoly31rt}	\\
&\hspace{-1mm} A_{3_1^r}^{-\infty}(x,y,a)/(a x^2 - 1)^2 = (a x - 1) y + a x(x-1)	\,	,	\label{apoly31ri}
\end{align}	
and
\begin{align}
B_{3_1^r}(a,b,x) &= x^2 (a-1)(a x-1) b^2 - \big(2 x^2 a^2 - x(x+1)a + 1 \big) b + x a^2 	\,	,	\label{bpoly31rt}	\\
B_{3_1^r}^{-\infty}(a,b,x) &= (a-1)(a x-1) b - q^2 x a^2 	\,	,	\label{bpoly31ri}
\end{align}	
respectively.
\eqref{apoly31rt} and \eqref{bpoly31rt} agree with the classical $a$-deformed $A$-polynomial and $x$-deformed $B$-polynomial in \eqref{aapoly31rtw} and \eqref{xbpoly31rtw}, respectively.
Also, \eqref{apoly31ri} and \eqref{bpoly31ri} agree with the first factors on the right-hand sides of \eqref{aapoly31rtw} and \eqref{xbpoly31rtw}, respectively, which arise from the formal solution $z=0$ of \eqref{svac31r}.

As in the case of $G=SU(2)$ \cite{Chung-ab}, applying $\hat{A}_{3_1^r}(\hat{x},\hat{y},a,q)$ to the homological block $\eqref{hb31ra}$ leads to an inhomogeneous equation, and the inhomogeneous term is annihilated by the operator $\hat{A}_{3_1^r}^{-\infty}(\hat{x},\hat{y},a,q)$ in \eqref{qapoly31ri}, which is associated with the formal solution $z=0$.
The homological block \eqref{hb31ra} is obtained by choosing a contour enclosing the poles $z=q^k$, $k=0,1, \ldots$, and extending toward $z=0$ in the $z$-plane, or $\log z= -\infty$ in the $\log z$-plane.
The asymptotic behavior of the integrand \eqref{int31r} as $z \rightarrow 0$ is given by $\exp \frac{1}{\hbar} \big( - \text{Li}_2(x) - \text{Li}_2(x^{-1} a^{-1}) + \text{Li}_2(a) +\log(-1) \log a + \frac{1}{2} (\log a)^2 \big)$, and the classical limits of the operators that annihilate it at order $\hbar^{-1}$ in the exponent are $y+x$ up to an overall factor and $(a-1)^2 b-a^2$, both of which are associated with the formal solution $z=0$.
We also note that the asymptotic behavior is the same as that of the inhomogeneous term.
Thus, it is natural that the homological block \eqref{hb31ra} is annihilated by $\hat{A}^{-\infty}_{3_1^r}(\hat{x},\hat{y},a,q) \hat{A}_{3_1^r}(\hat{x},\hat{y},a,q)$.	\\

When $N \in \mathbb{Z}_{\geq 2}$, acting with the operator $\hat{A}_{3_1^r}^{\text{nab}}(N;\hat{x},\hat{y},q)$, which corresponds to the non-abelian branch, on \eqref{hb31r} also yields the inhomogeneous equation, 
\begin{align}
\hat{A}_{3_1^r}^{\text{nab}}(N;\hat{x},\hat{y},q) F_{3_1^r}^{SU(N)}(x,q) = Z^{\text{inh-ab}}_{3_1^r}(N;x,q) + Z^{\text{inh-inf}}_{3_1^r}(N;x,q)	\,	.
\label{31rninh}
\end{align}
The sum of the inhomogeneous terms, which are, respectively, associated with the abelian branch and the formal solution $z=0$, is annihilated by the product $\hat{A}_{3_1^r}^{\text{inh}}(N;\hat{x},\hat{y},q) \hat{A}_{3_1^r}^{\text{ab}}(N;\hat{x},\hat{y},q)$ of two additional operators, where $\hat{A}_{3_1^r}^{\text{ab}}(N;\hat{x},\hat{y},q) Z^{\text{inh-ab}}_{3_1^r}(N;x,q)=0$.
For example, when $N=3$,
\begin{align}
\hat{A}_{3_1^r}^{\text{nab}}(3;\hat{x},\hat{y},q)	&= q^3 \hat{x}^3 (q^2 \hat{x} - 1) (q \hat{x}^2 - \hat{x} + 1) \hat{y} + (\hat{x} - 1) (q^3 \hat{x}^2 - q \hat{x} + 1)	\,	,	\\
\hat{A}_{3_1^r}^{\text{ab}}(3;\hat{x},\hat{y},q)	&= (q^2 \hat{x}^2 - 1) \hat{y} - q^2(q^4 \hat{x}^2 - 1)	\,	,	\\
\begin{split}
\hat{A}_{3_1^r}^{\text{inh}}(3;\hat{x},\hat{y},q) 	&= (q \hat{x} + 1) (q^3 \hat{x}^2 - 1) (q^3 \hat{x}^2 - q \hat{x} + 1) \hat{y} \\
&\hspace{37mm}+ q^3 \hat{x} (q^3 \hat{x} + 1) (q^5 \hat{x}^2 - 1) (q^5 \hat{x}^2 - q^2 \hat{x} + 1)	\,	.
\end{split}
\end{align}
Their classical limits are
\begin{align}
A_{3_1^r}^{\text{nab}}(3;x,y)	&= (x-1) (x^2-x+1) (x^3 y+1)	\,	,	\\
A_{3_1^r}^{\text{ab}}(3;x,y)	&= (x^2-1) (y-1)	\,	,	\\
A_{3_1^r}^{\text{inh}}(3;x,y)	&= (x-1) (x+1)^2 (x^2-x+1) (y+x)	\,	,	
\end{align}
so indeed $A_{3_1^r}^{\text{nab}}$ and $A_{3_1^r}^{\text{ab}}$ correspond to the non-abelian and abelian branches, respectively, and $A_{3_1^r}^{\text{inh}}$ is associated with the formal solution $z=0$.
By checking with several values of $N$, we see that $\hat{A}_{3_1^r}^{\text{ab}}(N;\hat{x},\hat{y},q)\hat{A}_{3_1^r}^{\text{nab}}(N;\hat{x},\hat{y},q)$ agrees with $\hat{A}_{3_1^r}(\hat{x},\hat{y},a,q)|_{a=q^N}$ up to an overall factor, and this should hold for general $N$.
Also, $\hat{A}_{3_1^r}^{\text{inh}}(3;\hat{x},\hat{y},q) (q^3 x^2-q x+1)$ agrees with $ (q^3 x^2-q x+1)(q^5 x^2-q^2 x+1) \hat{A}^{-\infty}_{3_1^r}(\hat{x},\hat{y},a,q)|_{a=q^3}$ up to an overall factor, and a similar relation should also hold for general $N$.	\\

Up to the inhomogeneous term, we also obtain the $\widehat{AB}_{3_1^r}$-ideal (or its subideal) generated by the Gr\"obner basis $\{ \hat{D}_{3_1^r} , \hat{A}_{3_1^r}\}$ with respect to the lexicographic order $\hat{a} \succ \hat{x}$, where
\begin{align}
\hat{D}_{3_1^r}(\hat{x}, \hat{a}, \hat{y}, \hat{b},q) = (\hat{a}-q) (\hat{a} \hat{x}^2 - q) \hat{b} + q \hat{a}^2 \hat{x} \hat{y} - q \hat{a}^2 \hat{x}	\,	.
\end{align}
The inhomogeneous term arises from the action of $\hat{A}_{3_1^r}$ in \eqref{qapoly31r}.
The annihilating ideal (or its subideal) for the inhomogeneous term is generated by $\{ \hat{A}_{3_1^r}^{-\infty}, \hat{B}_{3_1^r;A}^{-\infty} \}$ where
\begin{align}
\hat{B}_{3_1^r; A}^{-\infty}(\hat{a},\hat{b}, x, q) = (\hat{a}-q) (x\hat{a} - 1) (x^2 \hat{a} - q) \hat{b} + q^2 x (1 - q^2 x^2 \hat{a}) \hat{a}^2 	\,	.
\end{align}
For the lexicographic order $\hat{x} \succ \hat{a}$, the Gr\"obner basis is given by $\{ \hat{D}_{3_1^r} , \hat{B}_{3_1^r}\}$.
The inhomogeneous term arises from the action of $\hat{B}_{3_1^r}$ in \eqref{qbpoly31r}.
The annihilating ideal (or its subideal) for the inhomogeneous term is generated by $\{ \hat{B}_{3_1^r}^{-\infty} , \hat{A}_{3_1^r;B}^{-\infty} \}$ where
\begin{align}
\hat{A}_{3_1^r; B}^{-\infty}(\hat{x}, \hat{y}, a, q) = (a \hat{x}-1) \hat{y} + q a (q \hat{x}-1) \hat{x}	\,	.
\end{align}
For the homogeneous relations, the Gr\"obner bases are given by $\{ \hat{D}_{3_1^r} , \frac{1}{q (a q \hat{x}^2-1)} \hat{A}_{3_1^r}^{-\infty} \hat{A}_{3_1^r} \}$ and $\{ \hat{D}_{3_1^r} , \hat{B}_{3_1^r}^{-\infty} \hat{B}_{3_1^r}\}$ for the lexicographic orders $\hat{a} \succ \hat{x}$ and $\hat{x} \succ \hat{a}$, respectively.	\\

If taking the poles $z=xq^{k}$, $k=0,1, \ldots$, \eqref{int31r} gives
\begin{align}
\begin{split}
&\mathcal{B}_{3_1^r}(x,a,q) = \frac{\theta(q^3 a^{-1};q)}{\theta(q^2 x;q)}
\frac{\theta(q^{-1}ax;q)}{\theta(1;q)}
\frac{\theta(q^2 a^{-1}x^2;q)}{\theta(q^3 a^{-2} x;q)}
\frac{\theta(q^{-1}x;q)}{\theta(q^{-2} ax^2;q)} 	\\
&\hspace{40mm} \times \frac{(q^{-1}a;q)_\infty (q^{-1} ax^2;q)_\infty}{(x;q)_\infty (x^{-1};q)_\infty} 	
\sum_{k=0}^{\infty} \frac{q^{k} (q^{-1}ax;q)_k}{(q;q)_k (qx;q)_k (q^{-1} ax^2;q)_k}	\,	.
\end{split}
\label{31rnab}
\end{align}
It is annihilated by the $q$-difference operators \eqref{qapoly31rt} and \eqref{qbpoly31rt}.
Also, the Gr\"{o}bner bases obtained for \eqref{31rnab} are the same as those obtained above for the homological block $F_{3_1^r}(x,a,q)$ in the discussion of the $\widehat{AB}_{3_1^r}$-ideal.
When $a=q^N$ with $N \in \mathbb{Z}_{\geq 2}$, applying the operator $\hat{A}_{3_1^r}^{\text{nab}}(N;\hat{x}, \hat{y},q)$, discussed in \eqref{31rninh}, to \eqref{31rnab} also leads to an inhomogeneous equation.
The inhomogeneous term agrees with $Z_{3_1^r}^{\text{inh-inf}}$ in \eqref{31rninh} up to a sign, which we have checked for several values of $N$.
Denoting the annihilator for it by $\hat{A}^{-\infty}_{3_1^r}(N;\hat{x},\hat{y},q)$, \eqref{31rnab} is annihilated by $\hat{A}^{-\infty}_{3_1^r}(N;\hat{x},\hat{y},q) \hat{A}_{3_1^r}^{\text{nab}}(N;\hat{x}, \hat{y},q)$.
For example, when $N=3$, the operator $\hat{A}^{-\infty}_{3_1^r}(N;\hat{x},\hat{y},q)$ is given by 
\begin{align}
\hat{A}^{-\infty}_{3_1^r}(3;\hat{x},\hat{y},q) = (q \hat{x} + 1) \big(q^2 \hat{x}^2+ (q-2) q \hat{x} + 1\big) \hat{y} + q^3 \hat{x} (q^2 \hat{x}+1) \big(q^4 \hat{x}^2 + (q-2) q^2 \hat{x}+1 \big)	\,	.
\end{align}

The Weyl action $x \rightarrow q^2 a^{-1} x^{-1}$ on $\mathcal{B}_{3_1^r}(x,a,q)$ gives $\mathcal{B}_{3_1^r}(q^2 a^{-1} x^{-1},a,q)$.
The series expansion of $\mathcal{B}_{3_1^r}(q^2 a^{-1} x^{-1},a,q)$ in $x$ agrees with that obtained in \cite{EGGKPSS}.
$\mathcal{B}_{3_1^r}(x,a,q)$ and its Weyl partner are annihilated by the same $q$-difference operators \eqref{qapoly31rt} and \eqref{qbpoly31rt}, but $\mathcal{B}_{3_1^r}(x,a,q)$ doesn't agree with $\mathcal{B}_{3_1^r}(q^2 a^{-1} x^{-1},a,q)$ even when $a=q^N$, unlike the case of the left-handed trefoil knot.
Since the system has Weyl symmetry, the Weyl partner is expected to arise from the half-index calculation, though it is not obvious from \eqref{int31r} as in the case of the left-handed trefoil knot.	\\

As in the previous cases, we can also consider the integral,
\begin{align}
\begin{split}
&\hspace{-2mm}(q;q)_\infty \int \frac{dz}{2 \pi i z} \frac{1}{(z^{-1};q)_\infty} \frac{(q^{-1}a;q)_\infty}{(q^{-1}az;q)_\infty}
\frac{(qx^{-1};q)_\infty}{(q^{2}a^{-1}x^{-1}z^{-1};q)_\infty}
\frac{(q^{-1} ax;q)_\infty}{(z^{-1}x;q)_\infty}	\\
&\hspace{25mm} \times 
\frac{\theta((-q^{\frac{1}{2}})x^{-4} z;q)}{\theta((-q^{\frac{1}{2}})^{-3} a x^{4};q)} 
\frac{\theta((-q^{\frac{1}{2}})^{-1} x^2 z;q)}{\theta((-q^{\frac{1}{2}})^{-1} x^2;q)} 
\frac{\theta((-q^{\frac{1}{2}})^{-3} a x^{2} z;q)}{\theta((-q^{\frac{1}{2}}) a^{-1} x^{2};q)} 	\,	.
\end{split}
\label{int31ra-2}
\end{align}
The boundary conditions for the 3d chiral multiplets, $\Phi_{l=1, \ldots, 7}$, are also the same as in the case of \eqref{int41a-2} and \eqref{int31la-2}, while the 2d boundary multiplets in \eqref{int31ra-2} are different. 
The half-indices obtained from \eqref{int31ra-2} agree with those obtained from \eqref{int31r}, up to theta function ambiguity for generic $a$.
As in the case of the left-handed trefoil knot, there are two sets of poles $z=xq^k$ and $z=q^2 a^{-1}x^{-1}q^k$, $k=0,1, \ldots$, whose residues give $\mathcal{B}_{3_1^r}(x,a,q)$ and $\mathcal{B}_{3_1^r}(q^2 a^{-1}x^{-1},a,q)$, respectively, up to theta function ambiguities.
Therefore, for this choice of boundary conditions, the Weyl pair can be manifestly obtained from the residues at the two sets of poles.	\\

In addition, we find that, for generic $a$, the inhomogeneous parts arising from the action of $\hat{A}_{3_1^r}(\hat{x},\hat{y},a,q)$ on $\mathcal{B}_{3_1^r}(x,a,q)$ and on $\mathcal{B}_{3_1^r}(q^2 a^{-1} x^{-1},a,q)$ are the same, though $\mathcal{B}_{3_1^r}(x,a,q)$ and $\mathcal{B}_{3_1^r}(q^2 a^{-1} x^{-1},a,q)$ themselves don't agree.
Therefore, their difference is annihilated by $\hat{A}_{3_1^r}(\hat{x},\hat{y},a,q)$,
\begin{align}
\hat{A}_{3_1^r}(\hat{x},\hat{y},a,q) \big( \mathcal{B}_{3_1^r}(x,a,q) - \mathcal{B}_{3_1^r}(q^2 a^{-1} x^{-1},a,q) \big) = 0	\,	.
\end{align}
Also, when $a=q^N$, by checking for several values of $N$, we find that
\begin{align}
\hat{A}_{3_1^r}^{\text{nab}}(N;\hat{x}, \hat{y},q) \big( \mathcal{B}_{3_1^r}(x,q^N,q) - \mathcal{B}_{3_1^r}(q^{2-N}x^{-1},q^N,q) \big) = 0	\,	.
\end{align} 
Thus, their difference captures only the contribution from the non-abelian flat connection, without the common asymptotic part associated with the inhomogeneous term.
For example, when $N=2$, the difference is given by 
\begin{align}
(q;q)_\infty\frac{qx}{1-x} \frac{\theta((-q^{1/2})x^2;q)}{\theta((-q^{1/2})x;q)}	\,	.
\end{align}
Upon multiplying it by the unknot factor $x^{1/2}-x^{-1/2}$ and the overall factor $(-1)^{\log x / \log q}$, the resulting expression is annihilated by $\hat{y} - q^{-\frac{3}{2}}\hat{x}^{-3}$, which is the quantum $\hat{A}$-polynomial for the non-abelian branch in the case of the right-handed trefoil knot.
For reference, in the case of the left-handed trefoil knot, $\mathcal{B}_{3_1^l}(x,a,q)$ in \eqref{31lnab} at $a=q^2$ gives 
\begin{align}
(q;q)_\infty \frac{x}{q (x-1)} \frac{\theta((-q^{1/2})x^{-2};q)^2}{\theta((-q^{1/2})x^{-3};q) \theta((-q^{1/2})x^{-1};q)^2}	\,	.
\end{align}
Also, after multiplying it by $(-1)^{\log x / \log q}(x^{1/2}-x^{-1/2})$, the resulting expression is annihilated by $\hat{y} - q^{\frac{3}{2}}\hat{x}^{3}$, which is the quantum $\hat{A}$-polynomial for the non-abelian branch \cite{Beem-Dimofte-Pasquetti} in the case of the left-handed trefoil knot.
Thus, with a suitable normalization, including the full unknot factor for $G=SU(N)$ and the factor $(-1)^{\log x / \log q}$, the difference $\mathcal{B}_{3_1^r}(x,q^N,q) - \mathcal{B}_{3_1^r}(q^{2-N} x^{-1},q^N,q)$, which is the contribution solely from the non-abelian branch, would agree, up to theta function ambiguity, with the half-index obtained from the tetrahedra construction in \cite{Beem-Dimofte-Pasquetti, Dimofte-Gaiotto-Gukov, Dimofte-Gabella-Goncharov}, and similarly for $\mathcal{B}_{3_1^l}(x,q^N,q)$ or its Weyl partner.


\subsection{Critical point and contour}

Given the boundary conditions, a half-index is obtained by choosing a contour that passes through a critical point obtained from the supersymmetric vacuum equation \eqref{svac}. 
In the context of the 3d-3d correspondence, critical points correspond to flat connections of Chern-Simons theory.
When the parameter $a$ is turned on, both the abelian and non-abelian branches are captured in the effective twisted superpotential, or equivalently the $a$-deformed $A$-polynomial, whereas the abelian branch is lost at the level of the effective twisted superpotential when $a = 1$.
This can be seen from the $a$-deformed version of the critical point equation and the SUSY parameter equation, where we can track the critical points upon the limit $a \rightarrow 1$.	\\

As an example, we consider the case of the figure-eight knot.
When $a$ is turned on, the effective twisted superpotential \eqref{twspot41} and the critical point equation \eqref{svac} give
\begin{align}
1=-\frac{z (1-a z)}{(1-z) (1-zx^{-1}) (1-a x z)}	\,	.
\label{svac41}
\end{align}
In the semi-classical limit $q \rightarrow 1$, choosing $a = q^N$ with $N \in \mathbb{Z}_{\geq 2}$ yields $a \rightarrow 1$.
For generic values of $a$ with $a\neq 1$, \eqref{svac41} has three solutions, which are on equal footing and are not specifically associated with the abelian or non-abelian branches, as there is no such distinction for generic $a$.
When $a =1$, one of three solutions becomes $z=1$, and the factor $(1-z)$ in the numerator and the denominator cancels out, so the solution $z=1$ is not explicitly captured in the critical point equation.
Meanwhile, in the limit $a \rightarrow 1$, the SUSY parameter equation \eqref{spar} at $z=1$ leads to the abelian branch
\begin{align}
y = \frac{ (1-x) (1-zx^{-1})}{(1-a^{-1}x^{-1}) (1-a x z)} \bigg|_{a=1, \, z=1} = 1	\,	.
\label{abbr}
\end{align} 
When $a=1$, two other solutions of the supersymmetric vacuum equation \eqref{svac41} lead to the non-abelian branches of the $A$-polynomial.
If the homological block is expressed as the inverted Habiro series \eqref{hbn1}, in general, the SUSY parameter equation for $x$ and $y$ takes the same form as in \eqref{abbr} since the inverted Habiro coefficients don't depend on $x$, and the factor $(1-z)$ cancels out in the critical point equation \eqref{svac} when $a=1$.
Thus, when $a=1$, the solution $z=1$ corresponds to the abelian branch and is not captured at the level of the effective twisted superpotential.	\\

Therefore, when $a=q^N$, \textit{i.e.} $G=SU(N)$, the contour that passes through the critical point $z=1$ would be the contour that gives the homological block, which is the contribution from the abelian branch.
In the calculation of the half-index, the natural contour that passes through $z=1$ would be the contour enclosing the poles $z=q^k$, $k=0,1, \ldots$, from the $(z^{-1};q)_\infty^{-1}$ factor in the integrand.
More precisely, we may take a shift to the parameter $z \rightarrow (-q^{\frac{1}{2}})^{-1} z$ in the integrand as discussed in \cite{Chung-ab}.
The contour enclosing the poles $z=(-q^{\frac{1}{2}})q^k$, $k=0,1, \ldots$, gives the same homological block, and the critical point for the abelian branch is $z=-1$, so there is no overlap between the poles and the critical point.
Therefore, this would explain why the half-index obtained by taking the poles $z=q^k$, $k=0,1, \ldots$, from $(z^{-1};q)_\infty^{-1}$ gives the homological block.	\\

When $a$ is turned on, the branches are not distinguished as abelian or non-abelian but are instead on equal footing.
As can be seen in the examples discussed above, the half-indices obtained from the contours enclosing the different sets of poles have the same $a$-deformed quantum $\hat{A}$-polynomial.
It contains all branches, which at $a=1$ are distinguished as the abelian and non-abelian branches.


\subsection{Homological blocks for supergroups}

It is known that if the partition $\lambda$ satisfies the $L|M$-hook condition, $\lambda_{L+1} \leq M$, then the reduced $U_q(\mathfrak{gl}_{L|M})$ Reshetikhin-Turaev invariant of a knot colored by $\lambda$ is obtained from the reduced HOMFLY-PT polynomial colored by $\lambda$ at the specialization $a=q^{L-M}$ \cite{Queffelec2019}.
In particular, the one-row Young diagram $\lambda=(n-1)$ with $n-1$ boxes satisfy the hook condition when $L \geq 1$, since $\lambda_{L+1}=0$.
Therefore, the reduced $U_q(\mathfrak{gl}_{L|M})$ Reshetikhin-Turaev invariant of a knot colored by $\lambda=(n-1)$ is obtained from the reduced HOMFLY-PT polynomial colored by the totally symmetric representation with $n-1$ boxes at $a=q^{L-M}$ when $L \geq 1$.
Since we obtain the $a$-deformed homological block $F_K(x,a,q)$, or equivalently the $G=SU(N)$ homological block $F_K^{SU(N)}(x,q)$ at $a=q^N$, from the reduced HOMFLY-PT polynomial $P_K(n,a,q)$ for the totally symmetric representation with $n-1$ boxes, the homological block for $GL(L|M)$ colored by $\lambda=(n-1)$ would also be obtained from the $a$-deformed homological block at $a=q^{L-M}$.
For example, for the figure-eight and trefoil knots, they are given by
\begin{align}
&\hspace{-2mm} F_{4_1}^{GL(L|M)}(x,q) 	= \sum_{k=0}^{\infty} \frac{(-1)^k q^{\frac{1}{2}k(k+1)}  }{(x;q)_{k+L-M-1} (q^{2-L+M}x^{-1} ;q)_{k+L-M-1}} 
\frac{(q^{L-M-1};q)_k}{(q;q)_k}	\,	,
\label{hb41sg}	\\
&\hspace{-2mm} F_{3_1^l}^{GL(L|M)}(x,q) = 
-q^{-1} \sum_{k=0}^{\infty} \frac{q^{k^2} (-q^{\frac{1}{2}})^{-3(L-M-2)} q^{-\frac{1}{2}(L-M-2)^2} }{(x;q)_{k+L-M-1} (q^{2-L+M}x^{-1} ;q)_{k+L-M-1}} 
\frac{(q^{L-M-1};q)_k}{(q;q)_k}	\,	,
\label{hb31lsg}	\\
&\hspace{-2mm} F_{3_1^r}^{GL(L|M)}(x,q) = -q \sum_{k=0}^{\infty} \frac{q^{k} (-q^{\frac{1}{2}})^{3(L-M-2)} q^{\frac{1}{2}(L-M-2)^2}}{(x;q)_{k+L-M-1} (q^{2-L+M}x^{-1} ;q)_{k+L-M-1}} 
\frac{(q^{L-M-1};q)_k}{(q;q)_k}	\,	.
\label{hb31rsg}
\end{align}
The corresponding 3d $\mathcal{N}=2$ theories $T[S^3 \backslash K, GL(L|M)]$ can also be obtained in a similar way as described in the previous sections.


\section{Partition functions of $T[M_3]$}
\label{sec:pf}

Given the theories $T[M_3]$ obtained in the previous sections, we can compute the partition functions on some backgrounds.
In this section, we consider the partition functions on $S^2 \times_q S^1$ and $S_b^3$, and the twisted indices on $\mathcal{M}_{g,p}$, a degree-$p$ $S^1$-bundle over a genus-$g$ Riemann surface.\footnote{The partition function of $T[M_3]$ on the lens space $L(k,1)_b$, which corresponds to the complex Chern-Simons partition function at levels $k \in \mathbb{Z}$ and $\sigma = k \frac{1-b^2}{1+b^2} \in \mathbb{R}$, can also be computed from the appropriate finite sum and integral of the fused integrand, as in \eqref{pfti}, or from the sum of the fusions of half- and anti-half-indices, with appropriate assignments of the parameters \cite{Dimofte-lens}.}
In the 3d-3d correspondence, partition functions of $T[M_3]$ on $S^2 \times_q S^1$ and $S_b^3$ are matched with the complex Chern-Simons partition functions at integer levels 0 and 1, respectively \cite{Dimofte-Gaiotto-Gukov, Dimofte-Gaiotto-Gukov-index}.
The topologically twisted index on $\mathcal{M}_{g,p}$ is determined by the modular data of the MTC[$M_3$] for a 3-manifold $M_3$ \cite{Gukov-Putrov-Vafa}.


\subsection{Partition functions and twisted indices}
\label{ssec:pfti}

When the parameter $a$ is turned on, the partition functions on $S^2 \times_q S^1$ and $S_b^3$, and the twisted index on $S^2 \times_q S^1$ are obtained from 
\begin{align}
\hspace{-2mm}\sum_{m} \oint \frac{d\zeta}{2 \pi i \zeta} \big\| \Upsilon(z,x,a,q) \big\|^2_I		\,	,	\quad
\int_{\mathbb{R}} d\mu \big\| \Upsilon(z,x,a,q) \big\|^2_S		\,	,	\quad
\sum_{m} \int_{\text{JK}} \frac{d \zeta}{2 \pi i \zeta} \big\| \Upsilon(z,x,a,q) \big\|^2_A	\,	,
\label{pfti}
\end{align}
respectively.
Here, $\| \Upsilon \|^2 = \Upsilon(z,x,a,q) \Upsilon(\tilde{z},\tilde{x},\tilde{a},\tilde{q})$ where the prescription $(\tilde{u};\tilde{q})_\infty \leadsto (\tilde{q}^{-1} \tilde{u};\tilde{q}^{-1})_\infty^{-1}$ for $|\tilde{q}|>1$ is used to obtain $\Upsilon(\tilde{z},\tilde{x},\tilde{a},\tilde{q})$ from $\Upsilon(z,x,a,q)$.
The subscripts $I$, $S$, and $A$ denote the identity fusion, $S$-fusion, and $A$-twist fusion of the half- and anti-half-indices to build the partition functions on $S^2 \times_q S^1$ and $S_b^3$, and the twisted index on $S^2 \times_q S^1$, respectively \cite{Beem-Dimofte-Pasquetti, Benini-Zaffaroni, Nieri-Pasquetti, Gukov-Pei-Putrov-Vafa}.

The parameters depend on the fusion under consideration.
For the $S^2 \times_q S^1$ partition function, the parameters are given by 
\begin{align}
q= \tilde{q}^{-1} =: \bar{q}	\,	,	\quad	
z, \tilde{z}= q^{\frac{m}{2}}\zeta^{\pm 1}	\,	,	\quad	
x, \tilde{x}= q^{\frac{m_x}{2}} \zeta_x^{\pm 1}	\,	,	\quad	
a, \tilde{a} = q^{\frac{m_a}{2}} \zeta_a^{\pm 1}	\,	,
\end{align}
where $m$, $m_x$, and $m_a$, and $\zeta$, $\zeta_x$, and $\zeta_a$ are the magnetic fluxes and the fugacities for $U(1)_z$, $U(1)_x$, and $U(1)_a$, respectively.

For the $S_b^3$ partition function, 
\begin{align}
q=e^{2\pi i b^2}	\,	,	\
\bar{q} := \tilde{q}^{-1} = e^{-2\pi i b^{-2}}	\,	,	\quad	
z, \tilde{z} = e^{2 \pi b^{\pm 1} \mu}	\,	,	\
x, \tilde{x} = e^{2\pi b^{\pm 1} \mu_x}	\,	,	\
a, \tilde{a} = e^{2\pi b^{\pm 1} \mu_a}	\,	,
\end{align}
where $\mu$, $\mu_x$, and $\mu_a$ are the complexified mass parameters and $b$ is the squashing parameter of the 3-sphere.
In particular, the $S_b^3$ partition function can also be expressed in terms of the Faddeev quantum dilogarithm $\Phi_b(\mu)$, which is given by
\begin{align}
\Phi_b(\mu) = \frac{(-q^{\frac{1}{2}} z;q )_\infty}{(-\bar{q}^{\frac{1}{2}} \tilde{z}; \bar{q} )_\infty}	\,		\quad	\text{for } \text{Im} \, b^2 >0	\,	.
\end{align}
For example, the $S_b^3$ partition function for the figure-eight knot is given by
\begin{align}
\frac{\Phi_{b} (\mu_a - 3Q)}{\Phi_{b}(\mu_x - Q) \Phi_b(-\mu_a - \mu_x + 3Q)} \int_{\mathbb{R}- i \varepsilon} d\mu \frac{\Phi_b(\mu+\mu_x+\mu_a-3Q) \Phi_b(\mu - \mu_x + Q)}{\Phi_b(-\mu-Q) \Phi_b(\mu+\mu_a-3Q)}	\,	,
\end{align}
where $Q=\frac{i}{2}(b+b^{-1})$, and this provides an $a$-deformed state integral for the figure-eight knot.

For the twisted-index on $S^2 \times_q S^1$, the parameters are given by\footnote{With this identification of the parameters, we have $\big\| ((-q^{\frac{1}{2}})^{2-R} z^{-1} ;q)_\infty \big\|^2_{A} = ((-q^{\frac{1}{2}})^R q^{-\frac{m}{2}} \zeta;q)_\infty^{-1}$, $\big\| \theta((-q^{\frac{1}{2}})^{c} z;q) \big\|_{A}^{-2} = (-1)^{c^2} \zeta^{m} \big( (-1)^{m} \zeta^{-1} \big)^c$, and $\frac{\| \theta(z_1;q) \|_{A}^2 \| \theta(z_2;q) \|_{A}^2}{\| \theta(z_1 z_2;q) \|_{A}^2} = \zeta_1^{m_2} \zeta_2^{m_1}$, which are consistent with \cite{Benini-Zaffaroni, Closset:2019hyt}.} 
\begin{align}
q= \tilde{q}^{-1} =: \bar{q}	\,	,	\quad	
z, \tilde{z}= q^{\mp\frac{m}{2}}\zeta^{-1}	\,	,	\quad	
x, \tilde{x}= q^{\mp \frac{l}{2}} \xi^{-1}	\,	,	\quad	
a, \tilde{a} = q^{\mp \frac{s}{2}} \alpha^{-1}	\,	.
\end{align}
where $m$, $l$, and $s$, and $\zeta$, $\xi$, and $\alpha$ are the magnetic fluxes and the fugacities for $U(1)_z$, $U(1)_x$, and $U(1)_a$, respectively.

We first consider the partition functions for the figure-eight knot and take $T[S^3 \backslash 4_1]$ whose half-index is given by \eqref{int41a} with \eqref{integ41a}.
Taking the range $|q^2 \zeta_a^{-1}| < |\zeta_x| < 1 < |q\zeta_a^{-1}|$ for the $S^2 \times_q S^1$ partition function, and equivalently $\text{Re} [-b \mu_a + 4bQ] < \text{Re} [b \mu_x] < 0 < \text{Re} [-b \mu_a + 2bQ]$ for the $S_b^3$ partition function, there are three sets of poles inside the unit circle $|z|=1$ and in the upper half of the complex $\mu$-plane in \eqref{pfti}, respectively.
We take the contour to capture these poles.
For the twisted index on $S^2 \times_q S^1$, we take poles from chiral multiplets with negative charges under the $U(1)$ gauge group. 
There are no contributions from the boundaries at $\zeta=0,\infty$.
Then the integrals in \eqref{pfti} give the factorized forms
\begin{align}
Z_{4_1} = \| \mathcal{B}_{4_1,0}(x,a,q) \|^2_* + \| \mathcal{B}_{4_1,1}(x,a,q) \|^2_* + \|\mathcal{B}_{4_1,2}(x,a,q) \|^2_*
\label{indfact41}
\end{align}
where $*$ denotes either $I$, $S$ or $A$, corresponding to the identity, $S$-, and $A$-twist fusions, respectively.
Here,
\begin{align}
\begin{split}
\mathcal{B}_{4_1,0}(x,a,q)						&=
\frac{1}{(q;q)_\infty} \frac{(q x^{-1} ;q)_\infty (q^{-1} a x;q)_\infty}{ (x;q)_\infty (q^{2}x^{-1} a^{-1};q)_\infty}
\sum_{k=0}^{\infty} \frac{(-1)^k q^{\frac{1}{2}k(k+1)} (q^{-1}a;q)_k}{(q;q)_k (q x^{-1} ;q)_k (q^{-1} a x;q)_k}	\\
&= F_{4_1}(x,a,q)/(q;q)_\infty	\,	,
\end{split}	\label{fact410}	\\
\mathcal{B}_{4_1,1}(x,a,q)						&= 
\frac{(q;q)_\infty}{\theta((-q^{\frac{1}{2}})^3 a ^{-1}x^{-1};q)}\frac{(q^{-1}a x^2;q)_\infty (q^{-1}a;q)_\infty}{(x^{-1};q)_\infty (x;q)_\infty }
\sum_{k=0}^{\infty} \frac{(-1)^k x^k q^{\frac{1}{2}k(k+1)} (q^{-1} a x;q)_k}{(q;q)_k  (qx;q)_k (q^{-1} a x^2;q)_k}	\,	,
\label{fact411} 	\\
\mathcal{B}_{4_1,2}(x,a,q) 						&= 
\frac{(q;q)_\infty}{\theta((-q^{\frac{1}{2}})x^{-1};q)} \frac{(q^{3}a^{-1}x^{-2};q)_\infty (q^{-1}a;q)_\infty}{(q^{-2}ax;q)_\infty (q^2 a^{-1} x^{-1};q)_\infty}
\sum_{k=0}^{\infty} \frac{(-1)^k a^{-k} x^{-k}  q^{\frac{1}{2}k(k+5) }(q x^{-1};q)_k}{(q;q)_k  (q^3 a^{-1} x^{-1};q)_k (q^{3} a^{-1} x^{-2};q)_k}	\nonumber	\\
&=\mathcal{B}_{4_1,1}(q^2a^{-1}x^{-1},a,q) 	\,	,
\label{fact412}	
\end{align}
where $\mathcal{B}_{4_1,1}(x,a,q)$ and $\mathcal{B}_{4_1,2}(x,a,q)$ agree with the half-indices \eqref{41nab1} and \eqref{41nab2} up to theta function ambiguity and a $q$-dependent factor.

If we simply take poles $\zeta = \zeta_a^{-1} q^{1-\frac{|m+m_a|}{2}-j}$ with $0 \leq \frac{1}{2} \big(|m+m_a| \pm (m+m_a) \big)+j \leq n-1$ for the $S^2 \times_q S^1$ partition function, $\mu = - \mu_a + 2 Q -i h b-i \tilde{h}b^{-1}$ with $h, \tilde{h} =0,1, \ldots, n-1$ for the $S_b^3$ partition function, or $\zeta = \alpha q^{-1-\frac{1}{2}(m+s)+j}$ with $0 \leq j, \, (m+s+2)-j \leq n-1$ for the $S^2 \times_q S^1$ twisted index, and set $a=q^N$ and $x=q^n$, each partition function \eqref{pfti} is given by the fusion $Z_{4_1,\text{poly}} = \| P_{4_1}(n, a, q) \|_*^2$ of the colored HOMFLY-PT polynomial \eqref{homfly41}.	\\

By performing the similar calculations, the partition functions for the left-handed trefoil knot are given by
\begin{align}
Z_{3_1^l} = \| \mathcal{B}_{3_1^l,0}(x,a,q) \|^2_* + \| \mathcal{B}_{3_1^l,1}(x,a,q) \|^2_* +  \| \mathcal{B}_{3_1^l,1}(q^2a^{-1}x^{-1},a,q) \|^2_*
\label{indfact31l}
\end{align}
where
\begin{align}
\mathcal{B}_{3_1^l,0}(x,a,q)			&= 
 \frac{a^{\frac{1}{2}}\theta(a;q)}{(q;q)_\infty}
\frac{(q^{-1}ax;q)_\infty (qx^{-1};q)_\infty}{(x;q)_\infty (q^2 a^{-1} x^{-1};q)_\infty} 
\sum_{k=0}^{\infty} \frac{q^{k^2}  (q^{-1}a;q)_k}{(q;q)_k (q x^{-1};q)_k (q^{-1}a x ;q)_k }	\,	,	\label{fact31l0}	\\
\begin{split}
\mathcal{B}_{3_1^l,1}(x,a,q)			&= 
 \frac{a^{\frac{1}{2}} x^{-\frac{1}{2}}\theta(a;q)(q;q)_\infty }{\theta(x;q) \theta((-q^{\frac{1}{2}})^{-3} a x;q)} \frac{(q^{-1} a x^2;q)_\infty (q^{-1} a;q)_\infty }{(x;q)_\infty  (x^{-1};q)_\infty} 	\\
&\hspace{60mm}\times \sum_{k=0}^{\infty} \frac{x^{2k} q^{k^2} (q^{-1} a x;q)_k}{(q;q)_k  (qx;q)_k (q^{-1} a x^2;q)_k} 	\,	.
\label{fact31l1}
\end{split}
\end{align}
Up to theta function ambiguity and $q$-dependent overall factors, $\mathcal{B}_{3_1^l,j=0,1}$ agree with the half-indices \eqref{hb31la} and \eqref{31lnab} for the left-handed trefoil knot.
In particular, when taking $a=q^N$, $\mathcal{B}_{3_1^l,0}$ is the same as the homological block \eqref{hb31l} up to an overall $q$-dependent factor.
In \eqref{indfact31l}, the third term $\| \mathcal{B}_{3_1^l,1}(q^2a^{-1}x^{-1},a,q) \|^2_*$ would be regarded as arising from the Weyl completion of $\| \mathcal{B}_{3_1^l,1}(x,a,q) \|^2_*$.	\\

The partition functions for the right-handed trefoil knot is given by
\begin{align}
Z_{3_1^r} = \| \mathcal{B}_{3_1^r,0}(x,a,q) \|^2_* + \| \mathcal{B}_{3_1^r,1}(x,a,q) \|^2_* +  \| \mathcal{B}_{3_1^r,1}(q^2a^{-1}x^{-1},a,q) \|^2_*
\label{indfact31r}
\end{align}
where	
\begin{align}
\begin{split}
&\mathcal{B}_{3_1^r,0}(x,a,q)			= 
 \frac{a^{-\frac{1}{2}}\theta(a;q)^{-1}}{(q;q)_\infty}
\frac{(q^{-1}ax;q)_\infty (qx^{-1};q)_\infty}{(x;q)_\infty (q^2 a^{-1} x^{-1};q)_\infty} 	\\
&\hspace{70mm}\times \sum_{k=0}^{\infty} \frac{q^{k}  (q^{-1}a;q)_k}{(q;q)_k (q x^{-1};q)_k (q^{-1}a x ;q)_k }	\,	,
\end{split}	\label{fact31r0}	\\
\begin{split}
&\mathcal{B}_{3_1^r,1}(x,a,q)			= 
 \frac{a^{-\frac{1}{2}} x^{\frac{1}{2}}\theta(x;q)(q;q)_\infty }{\theta(a;q) \theta((-q^{\frac{1}{2}})^{-3} a x;q)} \frac{(q^{-1} a x^2;q)_\infty (q^{-1} a;q)_\infty }{(x;q)_\infty  (x^{-1};q)_\infty} 	\\
&\hspace{70mm}\times \sum_{k=0}^{\infty} \frac{q^{k} (q^{-1} a x;q)_k}{(q;q)_k  (qx;q)_k (q^{-1} a x^2;q)_k} 	\,	.
\label{fact31r1}
\end{split}
\end{align}
As in the previous examples, $\mathcal{B}_{3_1^r,j=0,1}$ agree with the half-indices \eqref{hb31ra} and \eqref{31rnab} for the right-handed trefoil knot, up to theta function ambiguity and $q$-dependent overall factors.
In particular, when $a=q^N$, $\mathcal{B}_{3_1^r,0}$ gives the homological block \eqref{hb31r} up to an overall $q$-dependent factor.
There is also an additional term $\| \mathcal{B}_{3_1^r,1}(q^2a^{-1}x^{-1},a,q) \|^2_*$ in \eqref{indfact31r}, which would be regarded as arising from the Weyl completion of $\| \mathcal{B}_{3_1^r,1}(x,a,q) \|^2_*$.	\\

For $G=SU(N)$ with $N \geq 2$, $\|(z^{-1};q)_\infty (q^{-1}az;q)_\infty \|^{-2}_*$ in the integrands \eqref{pfti} is given by
\begin{align}
\frac{(\bar{q} \tilde{z}^{-1};q)_\infty (\bar{q}^{2-N} \tilde{z};q)_\infty}{(z^{-1};q)_\infty (q^{N-1}z;q)_\infty} = \frac{\theta((-\bar{q}^{\frac{1}{2}})^{-1}\tilde{z};\bar{q})}{\theta((-q^{\frac{1}{2}})z;q)} (qz;q)_{N-2} (\bar{q}^{2-N}\tilde{z};\bar{q})_{N-2}	\,	,
\label{integfct}
\end{align}
where the set of poles of $\| (z^{-1};q)_\infty\|^{-2}_*$ is associated with the abelian branch.
The factor $\| \theta((-q^{\frac{1}{2}}) z ;q)\|^{-2}_*$ in \eqref{integfct} is $(-q^{\frac{1}{2}} \zeta)^m$, $(q^{-1}\bar{q})^{\frac{1}{24}} e^{-\pi i (\mu + Q)^2}$, and $(-\zeta)^{-(m+1)}$ for the $S^2 \times_q S^1$ partition function, $S_b^3$ partition function, and the $S^2 \times_q S^1$ twisted index, respectively.
Therefore, there is no set of poles associated to the abelian branch in the integral.
Thus, for $G=SU(N)$, the contribution from the abelian branch is absent in the partition functions, while the partition functions contain the contributions from the non-abelian branches.

For $G=SU(N)$, $\| \mathcal{B}_{K,j \neq 0}\|_*^2$ in \eqref{indfact41}, \eqref{indfact31l}, and \eqref{indfact31r}\footnote{Here and below, the notation $\| \mathcal{B}_{K,j \neq 0}\|_*^2$ for the trefoil knots also includes the last terms in \eqref{indfact31l} and \eqref{indfact31r}.} with $a=q^N$ and $\tilde{a}=\bar{q}^{-N}$ are the contributions from the non-abelian branches to the partition functions.
The chiral multiplet $\Phi_5$ has charges $(0,0,4-2N)$ under $U(1)_z \times U(1)_x \times U(1)_R$, and its contribution to $\| \mathcal{B}_{K,j \neq 0}\|_*^2$ with $a=q^N$ and $\tilde{a}=\bar{q}^{-N}$ is given by $\frac{(q^{N-1};q)_\infty}{(\bar{q}^{2-N} ;\bar{q})_\infty}$, which is divergent for $N \geq 2$.
Therefore, for the $G=SU(N)$ partition functions, we may decouple the contribution from the chiral multiplet $\Phi_5$. 
Then, the resulting partition functions are obtained by omitting the factor $\| (q^{-1}a;q)_\infty \|^2_*$ from all the terms $\| \mathcal{B}_{K,j}\|_*^2$ with $j\neq0$ and then taking the specialization of the sums of these terms at $a=q^N$ and $\tilde{a}=\bar{q}^{-N}$.\footnote{We further discuss the partition functions at and near $a=q^N$ in section \ref{aqn}.}


\subsubsection*{Topologically twisted index on $\mathcal{M}_{g,p}$}

The topologically twisted index on $\mathcal{M}_{g,p}$ is given by
\begin{align}
Z[\mathcal{M}_{g,p}] = \sum_{\beta} \mathcal{H}_\beta^{g-1} \mathcal{F}_\beta^{p}
\end{align}
where $\beta$ denotes the supersymmetric vacua, or the Bethe vacua \cite{NS-curved, Benini-Zaffaroni-2, Closset-Kim, Closset-Kim-Willett-1}.
The handle-gluing operator $\mathcal{H}_\beta$ and the fibering operator $\mathcal{F}_\beta$ are given by
\begin{align}
\mathcal{H}_\beta = e^{-2 \widetilde{\mathcal{W}}^{(0)}} \det \bigg( - \frac{\partial^2 \widetilde{\mathcal{W}}}{\partial (\log z)^2} \bigg) \bigg|_{z=z_\beta}	\,	,	\quad	
\mathcal{F}_\beta = e^{\frac{1}{2\pi i} \big( - \widetilde{\mathcal{W}} + \log z \frac{\partial \widetilde{\mathcal{W}}}{\partial \log z} + \log u \frac{\partial \widetilde{\mathcal{W}}}{\partial \log u }\big)}\Big|_{z=z_\beta}
\end{align}
where $z$ and $u$ denote the fugacities for the gauge and global symmetries \cite{Closset-Kim-Willett-1, Gang:2019jut}.\footnote{Here, we used the fugacity notation of the half-index.
More precisely, the fugacities $z$, $x$, and $a$ for the half-index should be replaced by the appropriate fugacities for the twisted index or the partition function.
For example, $z$, $x$, and $a$ are replaced by $\zeta^{-1}$, $\xi^{-1}$, and $\alpha^{-1}$ for the twisted index on $S^2 \times S^1$, and by $e^{2 \pi \mu}$, $e^{2\pi \mu_x}$, and $e^{2 \pi \mu_a}$ for the partition function on $S^3_{b=1}$, respectively.
}
The effective twisted superpotential $\widetilde{\mathcal{W}}:=\widetilde{\mathcal{W}}^{(-1)}$ is the leading order term of the perturbative expansion of the integrand of the half-index $\Upsilon \rightarrow \exp \big( \sum_{j=0} \hbar^{j-1} \widetilde{\mathcal{W}}^{(j-1)} \big)$, and $\widetilde{\mathcal{W}}^{(0)}$ is the next-to-leading-order term.
In the 3d-3d correspondence, the operators $\mathcal{H}_\beta$ and $\mathcal{F}_\beta$ are given by $S_{0\beta}^{-2}$ and $T_{\beta}$, respectively, where the $S$ and $T$ are modular matrices of $\text{MTC}[M_3]$ \cite{Gukov-Putrov-Vafa}.		\\

From the examples,\footnote{The handle-gluing and fibering operators of the twisted indices on $\mathcal{M}_{g,p}$ for the figure-eight knot and the left- and right-handed trefoil knots are given in Appendix A.} we can see that one of the solutions to the supersymmetric vacuum equation \eqref{svac} gives $\mathcal{H}_0 = \frac{1}{\Delta_K(x)^2}$ when $a =1$, where $\Delta_K(x)$ is the Alexander polynomial of a knot $K$, as expected.
Indeed, when $a=1$, \textit{i.e.} $N=0$, the $q \rightarrow 1$ limit of the homological block $F_K(x,a,q)$ is given by $\Delta_K(x)$.\footnote{In general, $\lim_{q\rightarrow 1} F_K(x,q^N,q) = \Delta_K(x)^{1-N}$ \cite{EGGKPS}.}
The contribution from the homological block to the twisted index on $S^2 \times S^1$ calculated from \eqref{pfti} is given by $Z^{\text{twisted}}_{K,0}[S^2 \times S^1] = \lim_{q \rightarrow 1} F_{K}(x,1,q)^2 = \Delta_K(x)^2$, which is equivalently given by $\mathcal{H}_0^{g-1}$ with $g=0$ for $S^2$.


\subsubsection*{Anti-half-index for $K$ and half-index for $m(K)$}

From the examples above, we see that the anti-holomorphic parts, which are given by the extensions of the half-indices $\mathcal{B}_{K,j}$ to $|\tilde{q}|>1$, are related to the half-indices $\mathcal{B}_{m(K),j}$ of a mirror knot $m(K)$.
More precisely, we find that
\begin{align}
\mathcal{B}_{K,j}(\tilde{x},\tilde{a},\tilde{q}) \simeq \mathcal{B}_{m(K),j}(\tilde{x}^{-1},\tilde{a}^{-1},\tilde{q}^{-1})	\,	,
\end{align}
up to theta function ambiguity and $q$-dependent factors.
Using this relation, the partition function can be expressed in terms of half-indices of $T[S^3 \backslash K]$ and $T[S^3 \backslash m(K)]$,
\begin{align}
Z_K \simeq \sum_j \mathcal{B}_{K,j}(x,a,q) \, \mathcal{B}_{m(K),j}(\tilde{x}^{-1},\tilde{a}^{-1},\tilde{q}^{-1})	\,	.	\label{pfkmk}
\end{align}
The anti-half-index $\mathcal{B}_{K,j}(\tilde{x},\tilde{a},\tilde{q})$ is the quantity that is fused with the half-index $\mathcal{B}_{K,j}(x,a,q)$ to form the partition functions.
The parameters $\tilde{x}$, $\tilde{a}$, and $\tilde{q}$ incorporate both the $SL(2,\mathbb{Z})$ action on the boundary torus of $D^2 \times_q S^1$ and the inversion due to orientation reversal in the gluing of the boundary tori \cite{Beem-Dimofte-Pasquetti}.
Meanwhile, $\mathcal{B}_{m(K),j}$ is also obtained by extending $\mathcal{B}_{K,j}$ via the inversion $q \rightarrow q^{-1}$.\footnote{Since $x$ and $a$ are analytic continuations of $q^n$ and $q^N$, respectively, it would be natural to also take $x \rightarrow x^{-1}$ and $a \rightarrow a^{-1}$ upon the extension.}
In a sense, $\mathcal{B}_{m(K),j}$ already incorporates the effect of orientation reversal in the context of the fusion of half- and anti-half-indices.
Therefore, the anti-half-index $\mathcal{B}_{K,j}(\tilde{x},\tilde{a},\tilde{q})$ is realized as $\mathcal{B}_{m(K),j}(\tilde{x}^{-1},\tilde{a}^{-1},\tilde{q}^{-1})$ without applying a further inversion to the parameters $\tilde{x}^{-1}$, $\tilde{a}^{-1}$, and $\tilde{q}^{-1}$.

For the $S^2 \times_q S^1$ partition function, when the magnetic fluxes for the global symmetries are turned off, the partition function \eqref{pfkmk} simplifies to
\begin{align}
Z_K[S^2 \times_q S^1](\zeta_x, \zeta_a, q) \simeq \sum_j \mathcal{B}_{K,j}(\zeta_x,\zeta_a,q) \mathcal{B}_{m(K),j}(\zeta_x,\zeta_a,q)	\,	.
\end{align}
In particular, the contribution from the homological block to the $S^2 \times_q S^1$ partition function is given by $Z_{K,0}[S^2 \times_q S^1](\zeta_x, \zeta_a, q) \simeq F_K(\zeta_x,\zeta_a,q) F_{m(K)}(\zeta_x,\zeta_a,q)$.


\subsection{Behavior of $a$-deformed partition functions at and near $a=q^N$}
\label{aqn}

We discuss the behavior of the partition function at and near $a = q^N$ with $N \in \mathbb{Z}_{\geq 2}$. 
When $a=q^N$, $\mathcal{B}_{K,j=0}$ and $\mathcal{B}_{K,j\neq0}$ correspond to the contributions from the abelian and non-abelian branches, respectively.
The condition $a=q^N$ can be realized, for example, by taking $(\zeta_a, m_a)=(q^{\frac{1}{2}(N + \tilde{N})}, N-\tilde{N})$ for the $S^2 \times_q S^1$ partition function, or $\mu_a=ibN+ib^{-1}\tilde{N}$ for the $S_b^3$ partition function, with $\tilde{N} \in \mathbb{Z}$, and this gives $\tilde{a} = \bar{q}^{-\tilde{N}}$.
The case $\tilde{N}=N$ corresponds to the partition function for $G=SU(N)$, which was also discussed in section \ref{ssec:pfti}.	\\

The case $(a,\tilde{a})=(q^N,\bar{q}^{-\tilde{N}})$ can be discussed in the context of Higgsing a global symmetry by giving a (space-dependent) vev to an operator in an appropriate background monopole sector, which is realized by taking the residue at a pole of the partition functions \cite{GRR-bootstrap, Chung-Dimofte-Gukov-Sulkowski}.
The term $\| \mathcal{B}_{K,0}(x,a,q) \|^2_*$ in the partition functions contains $(q^{-1}a;q)_k$ and $(\bar{q}^{-1}\tilde{a}^{-1};\bar{q})_k$ in the half-index and anti-half-index parts, respectively.
At $(a,\tilde{a})=(q^N,\bar{q}^{-\tilde{N}})$, $\| \mathcal{B}_{K,0}(x,a,q) \|^2_*$ is regular and simply gives $\| \mathcal{B}_{K,0}(x,a,q) \|^2_*\big|_{a=q^N, \tilde{a}=\bar{q}^{-\tilde{N}}}$, which is the contribution from the abelian branch, or from the homological block, when $N=\tilde{N}$.
The other parts of the partition functions, \textit{i.e.} $\| \mathcal{B}_{K,j\neq0}(x,a,q) \|^2_*$ in \eqref{indfact41}, \eqref{indfact31l}, and \eqref{indfact31r}, are the contributions from the non-abelian branches when $a=q^N$ and $\tilde{a}=\bar{q}^{-N}$, and they contain the factor $\frac{(q^{-1}a;q)_\infty}{(\bar{q}^2 \tilde{a};\bar{q})_\infty}$, which has a pole at $(a,\tilde{a})=(q^N,\bar{q}^{-\tilde{N}})$ when $N, \tilde{N} \geq 2$.
This pole is associated with the chiral operator $\Phi_5$ of charge $(-1,4)$ under $U(1)_a \times U(1)_R$.
Since we want to preserve the $U(1)_R$ symmetry under the Higgsing, we take $a'=q^{-2}a$ to be the fugacity for the global symmetry that we denote by $U(1)_{a'}$, as discussed in section \ref{ssec:hbhi} in the context of the $T[M_K]$ theory arising in the large-$N$ limit, rather than working with the fugacity $a$ for the $U(1)_a$ symmetry.\footnote{Thus, a natural context for discussing the Higgsing of the ``$U(1)_a$" symmetry would be the $T[M_K]$, where it gives rise to a vortex defect in the IR.
The $T[M_3]$ discussed in this work for $M_3=L_K$ or $M_K$, with the parameter $a$ or $a'$, arising from the large-$N$ limit, would admit another realization as a 3d defect theory coupled to a 5d $\mathcal{N}=1$ theory via geometric engineering \cite{Gadde-Gukov-Putrov-wall, Dimofte-Gukov-Hollands}.}  
Then, the factor $\frac{(q^{-1}a;q)_\infty}{(\bar{q}^2\tilde{a};\bar{q})_\infty}$ in the partition functions is expressed as $\frac{(q a';q)_\infty}{(\tilde{a}';\bar{q})_\infty}$, whose pole is associated with the operator $\Phi_5$, which now has charge $(-1,0)$ under $U(1)_{a'} \times U(1)_R$.
Taking a residue at the pole $(a',\tilde{a}')=(q^{N-2}, \bar{q}^{-(\tilde{N}-2)})$ in the partition functions can be understood as Higgsing the $U(1)_{a'}$ symmetry by giving a space-dependent vev to the operator $\Phi_5$ in the $(N-\tilde{N})$-th $U(1)_{a'}$ monopole sector.
In this way, the contribution from the non-abelian branches can survive but that from the abelian branch disappears.	\\

When $(a,\tilde{a})=(q^N,\bar{q}^{-\tilde{N}})$, $\| \mathcal{B}_{K,0}\|^2_*$ is regular, while the denominator of the factor $\frac{(q^{-1}a;q)_\infty}{(\bar{q}^2 \tilde{a};\bar{q})_\infty}$ in $\| \mathcal{B}_{K,j\neq0} \|^2_*$ is zero when $\tilde{N} \geq 2$ and nonzero when $\tilde{N} \leq 1$.
Thus, in the case $N = \tilde{N} \geq 2$, which corresponds to $G_\mathbb{C}=SL(N,\mathbb{C})$, $\| \mathcal{B}_{K,0} \|^2_*$ is finite and $\| \mathcal{B}_{K,j\neq0} \|^2_*$ is divergent.\footnote{We may also consider extending $N$ and $\tilde{N}$ to $N, \tilde{N} \leq 1$, including negative values, in the partition functions.
For example, when $N = \tilde{N} \leq 1$, $\| \mathcal{B}_{K,0} \|^2_*$ is non-trivial except for the case $N=\tilde{N}=1$, but the other parts $\| \mathcal{B}_{K,j\neq0} \|^2_*$ vanish.
It would be interesting to better understand the case $N=\tilde{N}\leq 0$, in particular in the context of supergroup Chern-Simons theory.}
Therefore, the ratio of the contribution from the abelian branch to that from the non-abelian branch is zero.
This is consistent with the general form of the partition function of the complex Chern-Simons theory on $M_3 = S^3 \backslash K$, which is expressed as a weighted sum over flat connections, with weights given by the inverse of the volume of the stabilizer, $\text{Stab}_\beta$, of the $SL(N,\mathbb{C})$ flat connection $\beta$,
\begin{align}
Z_{CS}[M_3] = \sum_{\text{flat } \beta} \frac{1}{\text{Vol}(\text{Stab}_\beta)} Z_\beta	\,	,
\label{csptfn}
\end{align}
where $\text{Vol}(\text{Stab}_\beta)$ is finite for irreducible flat connections and infinite for reducible flat connections \cite{Chung-Dimofte-Gukov-Sulkowski}.	\\

We may also take $a=q^{N+\epsilon}$ and study the limiting behavior of the partition functions as $\epsilon \rightarrow 0$.
For concreteness, we consider the $S^2 \times_q S^1$ partition function, which we express as $Z_{\text{ab}}+Z_{\text{nab}}$, where $Z_{\text{ab}} = \| \mathcal{B}_{K,j=0}(x,a,q) \|^2$ and $Z_{\text{nab}} = \sum_{j \neq 0} \| \mathcal{B}_{K,j}(x,a,q) \|^2$.
If we take $a=\tilde{a}^{-1}=q^{N+\epsilon}$, the $\epsilon$-series expansion of $Z_{\text{nab}}(x,\tilde{x},a,\tilde{a},q) = Z^{'}_{\text{nab}}(x,\tilde{x},a,\tilde{a},q) \frac{(q^{-1}a;q)_\infty}{(q^2 \tilde{a};q)_\infty}$ is given by
\begin{align}
\begin{split}
&\frac{(-1)^{N-2} q^{\frac{1}{2}(N-1)(N-2)}}{(q;q)_{N-2}^2}  \Bigg( \frac{Z^{'}_{\text{nab}}(x,\tilde{x},q^N,q^{-N},q)}{\log q} \epsilon^{-1}		\\
&\hspace{3mm}+\bigg(\frac{\partial}{\partial \log a} - \frac{\partial}{\partial \log \tilde{a}} + N- \frac{3}{2} -2 \sum_{j=N-1}^{\infty} \frac{q^j}{1-q^j} \bigg) Z^{'}_{\text{nab}}(x,\tilde{x},a,\tilde{a},q) \bigg|_{(a,\tilde{a})=(q^N,q^{-N})} + \mathcal{O}(\epsilon)\Bigg)	\,	.	
\end{split}
\end{align}
The contribution of the abelian branch is given by $Z_{\text{ab}}(x,\tilde{x},a,\tilde{a},q) \simeq Z_{\text{ab}}(x,\tilde{x},q^N,q^{-N},q) + \mathcal{O}(\epsilon)$.
Therefore, the leading order is given by the contribution from the non-abelian branch, while the contribution from the abelian branch appears in the next-to-leading order.

The deformation of $a=q^N$ discussed above and the general form \eqref{csptfn}, together with the resurgent analysis for the $G=SU(2)$ homological block for knot complements in \cite{Garoufalidis-Gu-Marino-Wheeler}, seem to indicate that, for $G=SU(N)$, the contribution $Z_{K,\text{ab}} = \mathcal{B}_{K,0}(x,q^N,q) \mathcal{B}_{K,0}(\tilde{x},\tilde{q}^N,\tilde{q})$ from the homological block wouldn't serve as a full partition function, whereas for the closed 3-manifolds the full partition function was conjectured to be given by the contributions from the homological blocks \cite{Gukov-Pei-Putrov-Vafa}.
It would be interesting to understand this better.


\acknowledgments{We would like to thank the Korea Institute for Advanced Study (KIAS) for hospitality during the final preparation of this manuscript.
This research was supported by the 2026 scientific promotion program funded by Jeju National University.
}


\begin{appendices}


\section{Twisted indices on $\mathcal{M}_{g,p}$ for $4_1$, $3_1^l$, and $3_1^r$ knots}
\label{sec:appendix}

In this appendix, we compute the handle-gluing and fibering operators of the twisted indices on $\mathcal{M}_{g,p}$ for the figure-eight knot and the left- and right-handed trefoil knots.

For the figure-eight knot, the effective twisted superpotential is given by \eqref{twspot41}, which we rewrite as
\begin{align}
\begin{split}
&-\widetilde{\mathcal{W}}_{4_1} = 
\widetilde{\mathcal{W}}_\Phi + \frac{1}{2} \Big( 2(\log a)^2 + 2 (\log x)^2 + 2 (\log z)^2 +2 \log a \log x + 2 \log a \log z \\
&\hspace{33mm} - 12 \log a \log(-1) - 8 \log x \log(-1) -4 \log z \log (-1) \\
&\hspace{93mm}+ 19 (\log (-1))^2 \Big)	- \frac{(2\pi i)^2}{12}	
\end{split}
\end{align}
where we added the contribution from the gravitational Chern-Simons term.
Here, $\widetilde{\mathcal{W}}_\Phi$ is
\begin{align}
\begin{split}
&\widetilde{\mathcal{W}}_\Phi = \text{Li}_2(z^{-1}) + \text{Li}_2(az) + \text{Li}_2(a^{-1}) + \text{Li}_2(x) \\
&\hspace{50mm}+ \text{Li}_2(a^{-1}x^{-1}z^{-1}) + \text{Li}_2(a^{-1}x^{-1}) + \text{Li}_2(xz^{-1})	\,	.
\end{split}
\end{align}
The next-to-leading order term $\widetilde{\mathcal{W}}^{(0)}$ is given by
\begin{align}
\begin{split}
&-2\widetilde{\mathcal{W}}_{4_1}^{(0)} = \log(1-z^{-1}) + 3 \log(1-az) - 3 \log(1-a^{-1}) + \log(1-x)	\\
&\hspace{20mm} - 3 \log(1-a^{-1}x^{-1}z^{-1}) + \log(1-xz^{-1}) -3 \log(1-a^{-1}x^{-1})	\\
&\hspace{20mm} - 6 \log a - 4 \log x - 2 \log z + 19 \log(-1)	\,	.
\end{split}
\end{align}
The solutions to the supersymmetric vacuum equation \eqref{svac41}, or the Bethe ansatz equation, 
\begin{align}
1=-\frac{z (1-a z)}{(1-z) (1-zx^{-1}) (1-a x z)}	\,	,
\end{align}
are given by $z_\beta(4_1;x,a) = \frac{\Lambda_\beta(x,a)}{3x}$, $\beta=0,1,2$, where 
\begin{align}
\begin{split}
&\Lambda_\beta(x,a) = a^{-1} \bigg( 2(ax^2+1)+ 2^{2/3} \omega^\beta (P+(P^2-4(ax^2-3ax+1)^3(ax^2+1)^3))^{1/3} \\
&\hspace{55mm} + \frac{2^{4/3} \omega^{-\beta} (ax^2-3ax+1)^3(ax^2+1)}{(P+(P^2-4(ax^2-3ax+1)^3(ax^2+1)^3))^{1/3}}	\bigg)
\end{split}
\end{align}
with $P(x,a) = (ax^2+1)(2ax^2-9ax+2)+27a^2 x^3$ and $\omega = \frac{1}{2} (-1+i \sqrt{3} )$.
Then, the handle-gluing operators $\mathcal{H}_\beta(4_1;x,a)$, $\beta=0,1,2$, are given by
\begin{align}
\mathcal{H}_\beta(4_1;x,a) = \frac{a^6 x (x-1) \Lambda_\beta^2 \Big(6-\frac{\Lambda_\beta}{x}\Big) \Big(6-\frac{\Lambda_\beta}{x^2} \Big) \Big(1-\frac{6 x}{a \Lambda_\beta} \Big)^3 \Big(-\frac{6 x}{a \Lambda_\beta- 6 x}-\frac{a \Lambda_\beta}{6-a \Lambda_\beta}+\frac{\Lambda_\beta}{\Lambda_\beta-6 x^2}+\frac{\Lambda_\beta}{\Lambda_\beta-6 x}-2 \Big)}{6 (1-a)^3 (6-a \Lambda_\beta)^3 (1-a x)^3}	\,	,
\end{align}
and the fibering operators $\mathcal{F}_\beta(4_1;x,a)$, $\beta=0,1,2$, are given by
\begin{align}
\mathcal{F}_\beta(K;x,a) = \exp \frac{1}{2\pi i} \Big(  \widetilde{\mathcal{W}}'_{\Phi - \partial \Phi}(z,x,a) + \widetilde{\mathcal{W}}'_{CS}(K;z,x,a) \Big) \bigg|_{z=z_\beta(K;x,a)}	\label{fbop}
\end{align}
where
\begin{align}
\begin{split}
&\widetilde{\mathcal{W}}'_{\Phi-\partial \Phi}(z,x,a) = \widetilde{\mathcal{W}}_{\Phi}(z,x,a)	 +\log z^{-1} \log (1-z^{-1})	 + \log (a z) \log(1-az)	\\
&\hspace{30mm}+ \log a^{-1} \log(1-a^{-1}) + \log x \log(1-x)		\\
&\hspace{30mm}+ \log (a^{-1}x^{-1}z^{-1}) \log(1-a^{-1}x^{-1}z^{-1}) \\
&\hspace{30mm}+ \log (xz^{-1}) \log(1-xz^{-1})+ \log (a^{-1}x^{-1}) \log(1-a^{-1}x^{-1})	\,	, 	\\
\end{split}	\\
\begin{split}
&\widetilde{\mathcal{W}}'_{CS}(4_1;z,x,a) =  -\frac{1}{2} \Big( 2(\log a)^2 + 2 (\log x)^2 + 2 (\log z)^2	\\
&\hspace{40mm} +2 \log a \log x + 2 \log a \log z - 19 (\log (-1))^2 \Big)
- \frac{(2\pi i)^2}{12}	\,	.
\end{split}
\end{align}

\bigskip

For the left-handed trefoil knot, the effective twisted superpotential can be expressed as
\begin{align}
\begin{split}
&-\widetilde{\mathcal{W}}_{3_1^l} = \widetilde{\mathcal{W}}_{\Phi}+ \frac{1}{2} \Big( 3(\log a)^2 + 2 (\log x)^2 + (\log z)^2 + 2 \log a \log x + 2 \log a \log z \\
&\hspace{33mm} - 14 \log a \log(-1) - 8 \log x \log(-1) - 2 \log z \log (-1)	\\
&\hspace{90mm} + 19 (\log (-1))^2 \Big)	- \frac{(2\pi i)^2}{12}	\,	,
\end{split}
\end{align}
and the next-to-leading order term is given by
\begin{align}
\begin{split}
&-2\widetilde{\mathcal{W}}_{3_1}^{(0)} = \log(1-z^{-1}) + 3 \log(1-az) - 3 \log(1-a^{-1}) + \log(1-x)	\\
&\hspace{20mm} - 3 \log(1-a^{-1}x^{-1}z^{-1}) + \log(1-xz^{-1}) -3 \log(1-a^{-1}x^{-1})	\\
&\hspace{20mm} - 7 \log a - 4 \log x - 1 \log z + 19 \log(-1)	\,	.
\end{split}
\end{align}
The solutions to the Bethe ansatz equation \eqref{svac31l} are given by
\begin{align}
z_{0}(3_1^l; x,a)	\,	,	\,	z_{1}(3_1^l; x,a) = \frac{a x^2+x+1 \mp S_{3_1^l}(x,a)}{2 \left(a x^2+a x-x+1\right)}	\,	,
\end{align}
respectively, where $S_{3_1^l}(x,a) = \sqrt{a^2 x^4-2 a x^3-2 a x^2+5 x^2-2 x+1}$.
The handle-gluing operators $\mathcal{H}_\beta(3_1^l;x,a)$, $\beta=0,1$, are given by
\begin{align}
\mathcal{H}_0(3_1^l; x,a)	\,	,	\,	\mathcal{H}_1(3_1^l; x,a) = \frac{a^2 x^2}{2(a-1)^3 (ax-1)^3} (\pm S_{3_1^l} U_{3_1^l} + S_{3_1^l}^2 V_{3_1^l})	\,	,
\end{align}
respectively, where
\begin{align}
\begin{split}
&\hspace{-2.5mm}U_{3_1^l}(x,a) = a^7 x^8-4 a^6 x^7-a^5 (4 a-13) x^6+a^4 (8 a-21) x^5-a^2 (4 a^2-13 a+18) x^3	\\
&\hspace{15mm} - a (a-2) (4 a^2-5 a+5)x^2+a^3 (6 a^2-21 a+28) x^4+(a-1)^3-x	\,	,
\end{split}	\\
\begin{split}
&\hspace{-2.5mm}V_{3_1^l}(x,a) = a^6 x^6-3 a^5 x^5+a^4 (8-3 a) x^4+a^3 (4 a-9) x^3+3 a^2 (a^2-3 a+3) x^2	\\
&\hspace{15mm}-a (a^2-3 a+3) x-(a-1)^3	\,	,
\end{split}
\end{align}
and the fibering operators $\mathcal{F}_\beta(3_1^l;x,a)$, $\beta=0,1$, are given by \eqref{fbop} with 
\begin{align}
\begin{split}
&\widetilde{\mathcal{W}}'_{CS}(3_1^l;z,x,a) = -\frac{1}{2} \Big( 3(\log a)^2 + 2 (\log x)^2 + (\log z)^2 \\
&\hspace{40mm}+ 2 \log a \log x + 2 \log a \log z - 19 (\log (-1))^2 \Big) - \frac{(2\pi i)^2}{12}	\,	.
\end{split}
\end{align}

\bigskip

Similarly, for the right-handed trefoil knot, the effective twisted superpotential is given by
\begin{align}
\begin{split}
&-\widetilde{\mathcal{W}}_{3_1^r} = \widetilde{\mathcal{W}}_\Phi + \frac{1}{2} \Big( (\log a)^2 + 2 (\log x)^2 + 3(\log z)^2 + 2 \log a \log x + 2 \log a \log z \\
&\hspace{33mm} - 10 \log a \log(-1) - 8 \log x \log(-1) - 6 \log z \log (-1)	\\ 
&\hspace{90mm}+ 19 (\log (-1))^2 \Big)	- \frac{(2\pi i)^2}{12}		\,	,
\end{split}
\end{align}
and the next-to-leading order term is given by
\begin{align}
\begin{split}
&-2\widetilde{\mathcal{W}}_{3_r}^{(0)} = \log(1-z^{-1}) + 3 \log(1-az) - 3 \log(1-a^{-1}) + \log(1-x)	\\
&\hspace{20mm} - 3 \log(1-a^{-1}x^{-1}z^{-1}) + \log(1-xz^{-1}) -3 \log(1-a^{-1}x^{-1})	\\
&\hspace{20mm} - 5 \log a - 4 \log x - 3 \log z + 19 \log(-1)	\,	.
\end{split}
\end{align}
The solutions to the Bethe ansatz equation \eqref{svac31r} are given by $z_{0}(3_1^r; x,a) = z_{1}(3_1^l; x,a)^{-1}$ and $z_{1}(3_1^r; x,a) = z_{0}(3_1^l; x,a)^{-1}$.
The handle-gluing operators $\mathcal{H}_\beta(3_1^r; x,a)$, $\beta=0,1$, are given by
\begin{align}
\mathcal{H}_0(3_1^r; x,a)	\,	,	\,	\mathcal{H}_1(3_1^r; x,a) = \frac{a^2 x^2}{2(a-1)^3 (ax-1)^3} (\mp S_{3_1^l} U_{3_1^l} + S_{3_1^l}^2 V_{3_1^l}) \bigg|_{x \rightarrow x^{-1},a \rightarrow a^{-1}}	\,	,
\end{align}
respectively, and the fibering operators $\mathcal{H}_\beta(3_1^r; x,a)$, $\beta=0,1$, are given by \eqref{fbop} with 
\begin{align}
\begin{split}
&\widetilde{\mathcal{W}}'_{CS}(3_1^r;z,x,a) = -\frac{1}{2} \Big( (\log a)^2 + 2 (\log x)^2 + 3(\log z)^2 \\
&\hspace{40mm}+ 2 \log a \log x + 2 \log a \log z - 19 (\log (-1))^2 \Big) - \frac{(2\pi i)^2}{12}	\,	.
\end{split}
\end{align}

\end{appendices}


\bibliographystyle{JHEP}
\bibliography{ref}

\providecommand{\href}[2]{#2}\begingroup\raggedright\begin{thebibliography}{10}

\bibitem{Dimofte-Gaiotto-Gukov}
T.~Dimofte, D.~Gaiotto and S.~Gukov, \emph{{Gauge Theories Labelled by
  Three-Manifolds}},
  \href{https://doi.org/10.1007/s00220-013-1863-2}{\emph{Commun. Math. Phys.}
  {\bfseries 325} (2014) 367--419},
  [\href{https://arxiv.org/abs/1108.4389}{{\ttfamily 1108.4389}}].

\bibitem{Dimofte-Gukov-Hollands}
T.~Dimofte, S.~Gukov and L.~Hollands, \emph{Vortex counting and lagrangian
  3-manifolds},  \href{https://arxiv.org/abs/1006.0977v1}{{\ttfamily
  1006.0977v1}}.

\bibitem{Terashima-Yamazaki}
Y.~Terashima and M.~Yamazaki, \emph{Sl(2,r) chern-simons, liouville, and gauge
  theory on duality walls},
  \href{https://arxiv.org/abs/1103.5748v1}{{\ttfamily 1103.5748v1}}.

\bibitem{Chung-Dimofte-Gukov-Sulkowski}
H.-J. Chung, T.~Dimofte, S.~Gukov and P.~Su\l{}kowski, \emph{{3d-3d
  Correspondence Revisited}},
  \href{https://doi.org/10.1007/JHEP04(2016)140}{\emph{JHEP} {\bfseries 04}
  (2016) 140}, [\href{https://arxiv.org/abs/1405.3663}{{\ttfamily 1405.3663}}].

\bibitem{Gukov-Putrov-Vafa}
S.~Gukov, P.~Putrov and C.~Vafa, \emph{{Fivebranes and 3-manifold homology}},
  \href{https://doi.org/10.1007/JHEP07(2017)071}{\emph{JHEP} {\bfseries 07}
  (2017) 071}, [\href{https://arxiv.org/abs/1602.05302}{{\ttfamily
  1602.05302}}].

\bibitem{Gukov-Pei-Putrov-Vafa}
S.~Gukov, D.~Pei, P.~Putrov and C.~Vafa, \emph{{BPS spectra and 3-manifold
  invariants}}, \href{https://doi.org/10.1142/S0218216520400039}{\emph{J. Knot
  Theor. Ramifications} {\bfseries 29} (2020) 2040003},
  [\href{https://arxiv.org/abs/1701.06567}{{\ttfamily 1701.06567}}].

\bibitem{Gukov-Manolescu}
S.~Gukov and C.~Manolescu, \emph{{A two-variable series for knot complements}},
  \href{https://doi.org/10.4171/qt/145}{\emph{Quantum Topol.} {\bfseries 12}
  (2021) 1--109}, [\href{https://arxiv.org/abs/1904.06057}{{\ttfamily
  1904.06057}}].

\bibitem{Gukov-Marino-Putrov}
S.~Gukov, M.~Marino and P.~Putrov, \emph{{Resurgence in complex Chern-Simons
  theory}},  \href{https://arxiv.org/abs/1605.07615}{{\ttfamily 1605.07615}}.

\bibitem{Chung-resurg}
H.-J. Chung, \emph{{Resurgent Analysis for Some 3-manifold Invariants}},
  \href{https://doi.org/10.1007/JHEP05(2021)106}{\emph{JHEP} {\bfseries 05}
  (2021) 106}, [\href{https://arxiv.org/abs/2008.02786}{{\ttfamily
  2008.02786}}].

\bibitem{GGP-walls}
A.~Gadde, S.~Gukov and P.~Putrov, \emph{Walls, lines, and spectral dualities in
  3d gauge theories},  \href{https://arxiv.org/abs/1302.0015v1}{{\ttfamily
  1302.0015v1}}.

\bibitem{Sugiyama-Yoshida}
Y.~Yoshida and K.~Sugiyama, \emph{{Localization of three-dimensional
  $\mathcal{N}=2$ supersymmetric theories on $S^1 \times D^2$}},
  \href{https://doi.org/10.1093/ptep/ptaa136}{\emph{PTEP} {\bfseries 2020}
  (2020) 113B02}, [\href{https://arxiv.org/abs/1409.6713}{{\ttfamily
  1409.6713}}].

\bibitem{Dimofte-Gaiotto-Paquette}
T.~Dimofte, D.~Gaiotto and N.~M. Paquette, \emph{{Dual boundary conditions in
  3d SCFT’s}}, \href{https://doi.org/10.1007/JHEP05(2018)060}{\emph{JHEP}
  {\bfseries 05} (2018) 060},
  [\href{https://arxiv.org/abs/1712.07654}{{\ttfamily 1712.07654}}].

\bibitem{Chung-3d3dpe}
H.-J. Chung, \emph{{3d-3d correspondence and 2d $\mathcal{N}$ = (0, 2) boundary
  conditions}}, \href{https://doi.org/10.1007/JHEP03(2024)085}{\emph{JHEP}
  {\bfseries 03} (2024) 085},
  [\href{https://arxiv.org/abs/2307.10125}{{\ttfamily 2307.10125}}].

\bibitem{Park-inverted}
S.~Park, \emph{{Inverted state sums, inverted Habiro series, and indefinite
  theta functions}},  \href{https://arxiv.org/abs/2106.03942}{{\ttfamily
  2106.03942}}.

\bibitem{Chung-ab}
H.-J. Chung, \emph{{3d-3d correspondence and abelian flat connection}},
  \href{https://arxiv.org/abs/2603.05236}{{\ttfamily 2603.05236}}.

\bibitem{Garoufalidis-Sun}
S.~Garoufalidis and X.~Sun, \emph{Thec–polynomial of a knot},
  \href{https://doi.org/10.2140/agt.2006.6.1623}{\emph{Algebraic \& Geometric
  Topology} {\bfseries 6} (Oct., 2006) 1623–1653}.

\bibitem{Mironov-Morozov-C}
A.~Mironov and A.~Morozov, \emph{{Algebra of quantum $ \mathcal{C}
  $-polynomials}}, \href{https://doi.org/10.1007/JHEP02(2021)142}{\emph{JHEP}
  {\bfseries 02} (2021) 142},
  [\href{https://arxiv.org/abs/2009.11641}{{\ttfamily 2009.11641}}].

\bibitem{EGGKPSS}
T.~Ekholm, A.~Gruen, S.~Gukov, P.~Kucharski, S.~Park, M.~Sto\v{s}i\'c et~al.,
  \emph{{Branches, quivers, and ideals for knot complements}},
  \href{https://doi.org/10.1016/j.geomphys.2022.104520}{\emph{J. Geom. Phys.}
  {\bfseries 177} (2022) 104520},
  [\href{https://arxiv.org/abs/2110.13768}{{\ttfamily 2110.13768}}].

\bibitem{EGGKPS}
T.~Ekholm, A.~Gruen, S.~Gukov, P.~Kucharski, S.~Park and P.~Su\l{}kowski,
  \emph{{${\widehat{Z}}$ at Large N: From Curve Counts to Quantum Modularity}},
  \href{https://doi.org/10.1007/s00220-022-04469-9}{\emph{Commun. Math. Phys.}
  {\bfseries 396} (2022) 143--186},
  [\href{https://arxiv.org/abs/2005.13349}{{\ttfamily 2005.13349}}].

\bibitem{Koutschan:holofunctions}
C.~Koutschan, \emph{{HolonomicFunctions} (user's guide)},  Tech. Rep. 10-01,
  RISC Report Series, Johannes Kepler University Linz, 2010.

\bibitem{FGS-superA}
H.~Fuji, S.~Gukov and P.~Sulkowski, \emph{{Super-A-polynomial for knots and BPS
  states}}, \href{https://doi.org/10.1016/j.nuclphysb.2012.10.005}{\emph{Nucl.
  Phys. B} {\bfseries 867} (2013) 506--546},
  [\href{https://arxiv.org/abs/1205.1515}{{\ttfamily 1205.1515}}].

\bibitem{Ooguri-Vafa}
H.~Ooguri and C.~Vafa, \emph{Knot invariants and topological strings},
  {\emph{Nucl. Phys.} {\bfseries B5777} (Jan, 2000) 419--438},
  [\href{https://arxiv.org/abs/hep-th/9912123v3}{{\ttfamily
  hep-th/9912123v3}}].

\bibitem{Witten-M5knots}
E.~Witten, \emph{Fivebranes and knots}, {\emph{Quantum Topol.} {\bfseries 3}
  (2012) 1--137}, [\href{https://arxiv.org/abs/1101.3216v1}{{\ttfamily
  1101.3216v1}}].

\bibitem{Beem-Dimofte-Pasquetti}
C.~Beem, T.~Dimofte and S.~Pasquetti, \emph{{Holomorphic Blocks in Three
  Dimensions}}, \href{https://doi.org/10.1007/JHEP12(2014)177}{\emph{JHEP}
  {\bfseries 12} (2014) 177},
  [\href{https://arxiv.org/abs/1211.1986}{{\ttfamily 1211.1986}}].

\bibitem{Aganagic-Vafa-Q}
M.~Aganagic and C.~Vafa, \emph{{Large N Duality, Mirror Symmetry, and a
  Q-deformed A-polynomial for Knots}},
  \href{https://arxiv.org/abs/1204.4709}{{\ttfamily 1204.4709}}.

\bibitem{CCFGH}
M.~C. Cheng, S.~Chun, F.~Ferrari, S.~Gukov and S.~M. Harrison, \emph{{3d
  Modularity}}, \href{https://doi.org/10.1007/JHEP10(2019)010}{\emph{JHEP}
  {\bfseries 10} (2019) 010},
  [\href{https://arxiv.org/abs/1809.10148}{{\ttfamily 1809.10148}}].

\bibitem{Dimofte-Gabella-Goncharov}
T.~Dimofte, M.~Gabella and A.~B. Goncharov, \emph{K-decompositions and 3d gauge
  theories},  \href{https://arxiv.org/abs/1301.0192v1}{{\ttfamily
  1301.0192v1}}.

\bibitem{Queffelec2019}
H.~Queffelec and A.~Sartori, \emph{A note on link invariants and the homfly--pt
  polynomial},  in \emph{Knots, Low-Dimensional Topology and Applications},
  pp.~279--294.
\newblock Springer International Publishing, 2019.
\newblock \href{https://doi.org/10.1007/978-3-030-16031-9_13}{DOI}.

\bibitem{Dimofte-lens}
T.~Dimofte, \emph{{Complex Chern{\textendash}Simons Theory at Level k via the
  3d{\textendash}3d Correspondence}},
  \href{https://doi.org/10.1007/s00220-015-2401-1}{\emph{Commun. Math. Phys.}
  {\bfseries 339} (2015) 619--662},
  [\href{https://arxiv.org/abs/1409.0857}{{\ttfamily 1409.0857}}].

\bibitem{Dimofte-Gaiotto-Gukov-index}
T.~Dimofte, D.~Gaiotto and S.~Gukov, \emph{{3-Manifolds and 3d Indices}},
  \href{https://doi.org/10.4310/ATMP.2013.v17.n5.a3}{\emph{Adv. Theor. Math.
  Phys.} {\bfseries 17} (2013) 975--1076},
  [\href{https://arxiv.org/abs/1112.5179}{{\ttfamily 1112.5179}}].

\bibitem{Benini-Zaffaroni}
F.~Benini and A.~Zaffaroni, \emph{{A topologically twisted index for
  three-dimensional supersymmetric theories}},
  \href{https://doi.org/10.1007/JHEP07(2015)127}{\emph{JHEP} {\bfseries 07}
  (2015) 127}, [\href{https://arxiv.org/abs/1504.03698}{{\ttfamily
  1504.03698}}].

\bibitem{Nieri-Pasquetti}
F.~Nieri and S.~Pasquetti, \emph{Factorisation and holomorphic blocks in 4d},
  \href{https://doi.org/10.1007/jhep11(2015)155}{\emph{Journal of High Energy
  Physics} {\bfseries 2015} (Nov., 2015) }.

\bibitem{Closset:2019hyt}
C.~Closset and H.~Kim, \emph{{Three-dimensional {\ensuremath{\mathscr{N}}} = 2
  supersymmetric gauge theories and partition functions on Seifert manifolds: A
  review}}, \href{https://doi.org/10.1142/S0217751X19300114}{\emph{Int. J. Mod.
  Phys. A} {\bfseries 34} (2019) 1930011},
  [\href{https://arxiv.org/abs/1908.08875}{{\ttfamily 1908.08875}}].

\bibitem{NS-curved}
N.~A. Nekrasov and S.~L. Shatashvili, \emph{{Bethe/Gauge correspondence on
  curved spaces}}, \href{https://doi.org/10.1007/JHEP01(2015)100}{\emph{JHEP}
  {\bfseries 01} (2015) 100},
  [\href{https://arxiv.org/abs/1405.6046}{{\ttfamily 1405.6046}}].

\bibitem{Benini-Zaffaroni-2}
F.~Benini and A.~Zaffaroni, \emph{{Supersymmetric partition functions on
  Riemann surfaces}}, {\emph{Proc. Symp. Pure Math.} {\bfseries 96} (2017)
  13--46}, [\href{https://arxiv.org/abs/1605.06120}{{\ttfamily 1605.06120}}].

\bibitem{Closset-Kim}
C.~Closset and H.~Kim, \emph{{Comments on twisted indices in 3d supersymmetric
  gauge theories}}, \href{https://doi.org/10.1007/JHEP08(2016)059}{\emph{JHEP}
  {\bfseries 08} (2016) 059},
  [\href{https://arxiv.org/abs/1605.06531}{{\ttfamily 1605.06531}}].

\bibitem{Closset-Kim-Willett-1}
C.~Closset, H.~Kim and B.~Willett, \emph{{Supersymmetric partition functions
  and the three-dimensional A-twist}},
  \href{https://doi.org/10.1007/JHEP03(2017)074}{\emph{JHEP} {\bfseries 03}
  (2017) 074}, [\href{https://arxiv.org/abs/1701.03171}{{\ttfamily
  1701.03171}}].

\bibitem{Gang:2019jut}
D.~Gang and M.~Yamazaki, \emph{{Expanding 3d $ \mathcal{N} $ = 2 theories
  around the round sphere}},
  \href{https://doi.org/10.1007/JHEP02(2020)102}{\emph{JHEP} {\bfseries 02}
  (2020) 102}, [\href{https://arxiv.org/abs/1912.09617}{{\ttfamily
  1912.09617}}].

\bibitem{GRR-bootstrap}
D.~Gaiotto, L.~Rastelli and S.~S. Razamat, \emph{Bootstrapping the
  superconformal index with surface defects},
  \href{https://arxiv.org/abs/1207.3577v1}{{\ttfamily 1207.3577v1}}.

\bibitem{Gadde-Gukov-Putrov-wall}
A.~Gadde, S.~Gukov and P.~Putrov, \emph{{Walls, Lines, and Spectral Dualities
  in 3d Gauge Theories}},
  \href{https://doi.org/10.1007/JHEP05(2014)047}{\emph{JHEP} {\bfseries 05}
  (2014) 047}, [\href{https://arxiv.org/abs/1302.0015}{{\ttfamily 1302.0015}}].

\bibitem{Garoufalidis-Gu-Marino-Wheeler}
S.~Garoufalidis, J.~Gu, M.~Marino and C.~Wheeler, \emph{{Resurgence of
  Chern-Simons theory at the trivial flat connection}},
  \href{https://arxiv.org/abs/2111.04763}{{\ttfamily 2111.04763}}.

\end{thebibliography}\endgroup

\end{document}